\documentclass{article}
    
        \usepackage[final]{neurips_2024}

    \usepackage[utf8]{inputenc} 
    \usepackage[T1]{fontenc}    
    \usepackage{hyperref}       
    \hypersetup{
    colorlinks=true,    
    linkcolor=blue,     
    citecolor=blue,     
    urlcolor=blue       
}
    \usepackage{url}            
    \usepackage{booktabs}       
    \usepackage{threeparttable}
    \usepackage{amsfonts}       
    \usepackage{nicefrac}       
    \usepackage{microtype}      
    \usepackage{xcolor}         
    \usepackage{graphicx}
    \usepackage{longtable}
    \usepackage{booktabs}
    \usepackage{caption}
    \usepackage{arydshln}
    \usepackage{amsmath}
    \usepackage{cleveref}
    \usepackage{subcaption}
    \usepackage{array}
    \usepackage{listings}
    \newif\ifshowcomments
    \showcommentstrue 
    
    \definecolor{alexcolor}{RGB}{0,100,0}      
    \definecolor{kunalcolor}{RGB}{0,0,139}     
    \definecolor{milescolor}{RGB}{139,0,0}     
    \definecolor{esincolor}{RGB}{148,0,211}    
    \definecolor{deepcolor}{RGB}{184,134,11}   
    \definecolor{saffroncolor}{RGB}{255,140,0} 
    \definecolor{lianecolor}{RGB}{105,42,42}   
    \definecolor{tedcolor}{RGB}{0,128,128}     
    
    \title{Clio: Privacy-Preserving Insights \\ into Real-World AI Use}

    \author{%
      Alex Tamkin\thanks{Equal Contribution},\; Miles McCain\footnotemark[1], \, Kunal Handa, Esin Durmus, Liane Lovitt, Ankur Rathi \\\\
      \textbf{Saffron Huang, Alfred Mountfield, Jerry Hong, Stuart Ritchie},\\ \textbf{Michael Stern, Brian Clarke, Landon Goldberg, Theodore R. Sumers},  \\ \textbf{Jared Mueller, William McEachen, Wes Mitchell, Shan Carter}\\\\
      \textbf{Jack Clark, Jared Kaplan, Deep Ganguli}\\\\
      Anthropic\\
    }

\begin{document}

    \maketitle

    \begin{abstract}

   How are AI assistants being used in the real world? While model providers in theory have a window into this impact via their users' data, both privacy concerns and practical challenges have made analyzing this data difficult. To address these issues, we present Clio (\textbf{Cl}aude \textbf{i}nsights and \textbf{o}bservations), a privacy-preserving platform that uses \textit{AI assistants themselves} to analyze and surface aggregated usage patterns across millions of conversations, without the need for human reviewers to read raw conversations. We validate this can be done with a high degree of accuracy and privacy by conducting extensive evaluations. We demonstrate Clio's usefulness in two broad ways. First, \textbf{we share insights about how models are being used in the real world} from one million Claude.ai Free and Pro conversations, ranging from \textit{providing advice on hairstyles} to \textit{providing guidance on Git operations and concepts}. We also identify the most common high-level use cases on Claude.ai (coding, writing, and research tasks) as well as patterns that differ across languages (e.g., conversations in Japanese discuss elder care and aging populations at higher-than-typical rates). Second, we use Clio to \textbf{make our systems safer} by identifying coordinated attempts to abuse our systems, monitoring for unknown unknowns during critical periods like launches of new capabilities or major world events, and improving our existing monitoring systems. We also discuss the limitations of our approach, as well as risks and ethical concerns. By enabling analysis of real-world AI usage, Clio provides a scalable platform for empirically grounded AI safety and governance.
    
    \end{abstract}

    \begin{figure}
        \centering
        \includegraphics[width=0.99\linewidth]{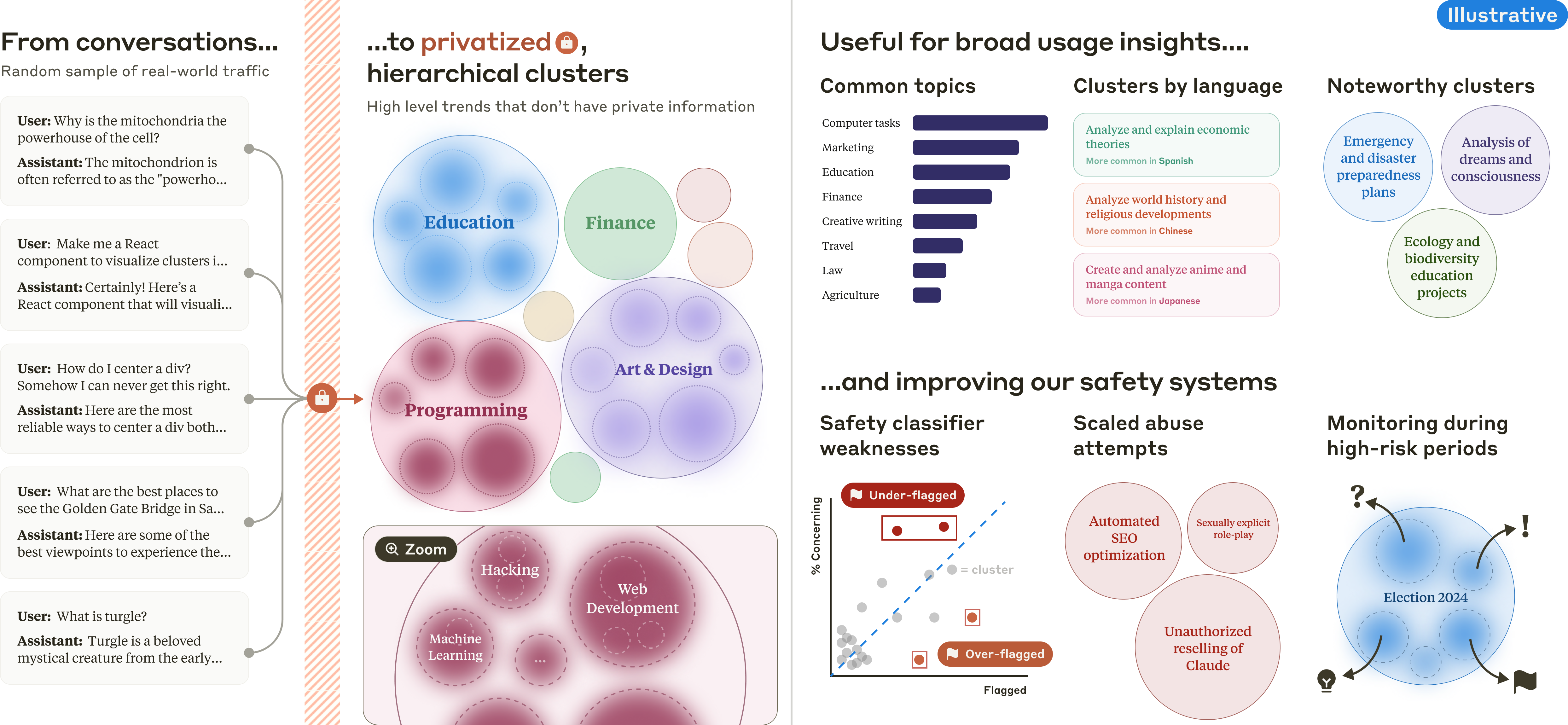}
        \caption{\textbf{Using Clio to understand real-world use of AI assistants.} Clio transforms raw conversations into high-level patterns and insights. This approach enables us to understand how AI assistants are being used in practice—analogous to how Google Trends provides insights about web search behavior. See \Cref{fig:system} for more details on how Clio works and how it preserves privacy. \textit{(Note: figure contains illustrative conversation examples only.})}
        \label{fig:clio-flow}
    \end{figure}
    
    \section{Introduction}
    \label{sec:intro}
    
    Despite widespread interest about the impact of AI systems on society, there is remarkably little public data about how models are actually being used in practice. What kinds of capabilities are seeing the most real-world adoption in the economy? How does usage vary across different communities and cultures? Which anticipated benefits and risks are most borne out in concrete data?
    
    This lack of understanding is particularly striking because model providers have access to \textit{usage} \textit{data} that could be used to answer these exact questions. Providers, however, face significant challenges in analyzing this data and sharing these potential insights:
    
     First, users share sensitive personal and business information with these systems, creating a fundamental tension between privacy protection and the need for providers to understand how their systems are being used. Second, having humans review conversations can raise ethical concerns, due to the repetitive nature of the task and the potentially distressing content reviewers could be exposed to. Third, providers face competitive pressures not to release usage data even if it would be in the public interest, as such data could reveal information about their user bases to competitors. And finally, the sheer scale of the data makes manual review impractical---millions of messages are sent daily, far more than any human team could meaningfully analyze. 
    
\begin{figure}
        \centering
        \includegraphics[width=0.99\linewidth]{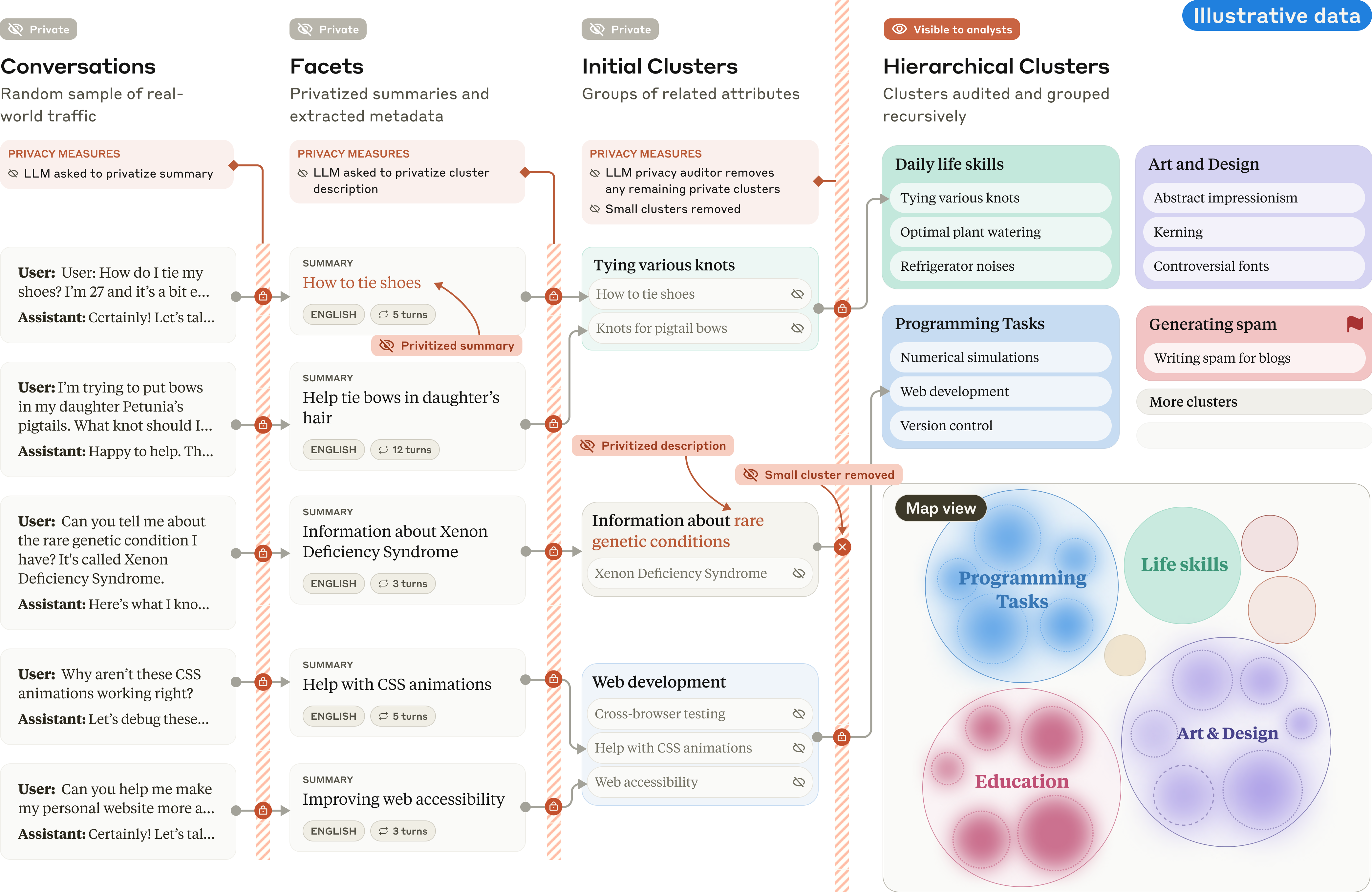}
        \caption{\textbf{System diagram.} This diagram illustrates how Clio processes insights from a sample of real-world conversations while maintaining user privacy. Clio processes a raw sample of traffic, extracts key facets (attributes like language or conversation topic), groups these facets into similar clusters (using text embeddings and k-means), and finally organizes those clusters into both a hierarchy as well as 2D space for ease of exploration. Along the way, Clio applies several privacy barriers (orange stripes) that prevent private information from reaching the user-visible parts of Clio (right). See \Cref{sec:system} for more details on each stage of the pipeline. (\textit{Note: figure contains illustrative examples only.})}
        \label{fig:system}
    \end{figure}

    To address these challenges, we present Clio (\textbf{Cl}aude \textbf{i}nsights and \textbf{o}bservations). Clio is a system that uses AI assistants themselves to surface aggregated insights across millions of model interactions while preserving user privacy (\Cref{fig:clio-flow,fig:system}).\footnote{Anthropic enforces strict internal privacy controls. Our \href{https://www.anthropic.com/legal/privacy}{privacy policy} enables us to analyze \textit{aggregated and anonymized} user interactions to understand patterns or trends. We continue to manage data according to our \href{https://privacy.anthropic.com/en/articles/10023548-how-long-do-you-store-personal-data}{privacy and retention policies}, and maintain our approach of not training our generative models on user conversations by default. Because we focus on studying patterns in individual usage, the results shared in this paper exclude activity from business customers (i.e. Team, Enterprise, and all API customers). For more information, see \Cref{sec:accesspolicies}.}  Similar to how Google Trends provides aggregate insights about web search behavior, Clio reveals patterns about how AI assistants are used in the real world. Clio then visualizes these patterns in an interface that enables discovering both specific patterns of interest as well as unknown unknowns (\Cref{fig:clio-interface}).
    
    First, we describe our \textit{evaluations} of Clio's outputs, including both the \textit{faithfulness} of the insights produced by Clio (\Cref{sec:validation}) as well as validating empirically that the final Clio outputs do not contain any private user data (\Cref{subsec:privacy-design} and \Cref{sec:privacy}). For example, Clio reconstructed the ground-truth distribution of topics in a synthetic dataset with 94\% accuracy, and produced no clusters with private data in an audit of 5,000 conversations (\Cref{sec:design}; \Cref{fig:supervised-reconstruction-regular,fig:privacy}). Next, we demonstrate Clio's capabilities and real-world impact across two broad use cases:
    
    \begin{enumerate}
        \item \textbf{Understanding broad usage patterns  (\Cref{sec:vignette-use}; \Cref{fig:claude_tasks,fig:multilingual_comparison}):} By clustering and visualizing millions of conversations, Clio reveals how people actually use AI assistants in the real world, without the need for human reviewers to read millions of user conversations. Specifically, we share the most common high-level use cases Clio identifies on Claude.ai Free and Pro. We find that coding and business use cases dominate, with significant activity in areas like debugging code, drafting professional emails, and analyzing business data. By comparing conversations across languages, we also find significant variations in how different linguistic communities use Claude. For example, Japanese and Chinese conversations are more likely to discuss elder care.
        \item \textbf{Improving Anthropic's safety systems  (\Cref{sec:vignette-safety}; \Cref{fig:ts-flagged-concerning-comparison}):} We share three case studies for how we have used Clio to enhance Anthropic's safety efforts: detecting coordinated misuse that is invisible at the individual conversation level, monitoring for unknown unknowns in periods of increased uncertainty such as the launch of new capabilities and in the run-up to major world events, and identifying patterns of over-triggering and under-triggering in our existing safety classifiers. Clio's insights have led to concrete enforcement actions on our systems; for example, we uncovered and banned a network of automated accounts that were attempting to abuse the free version of Claude.ai to generate spam optimized for search engines.
    \end{enumerate} 
    
    Finally, we discuss Clio's limitations (\Cref{sec:limitations}), and also discuss in detail the \textit{ethical considerations and potential risks} that Clio poses (\Cref{sec:risks-and-ethics}), along with mitigations. For example, we discuss the potential for Clio-like systems to be misused, additional potential privacy concerns, and more. Ultimately, we derive a justification for building and disclosing Clio that centers around the many pro-social and safety-enhancing features it affords both us and other model providers.
    
Looking ahead, the need for empirical understanding of how AI systems are used in practice will only increase as these systems become more capable and widespread. While pre-deployment testing like red-teaming \citep{ganguli2022red} and evaluations remain crucial, post-deployment monitoring \citep{stein2024role} provides an essential complement by surfacing real-world usage patterns and risks that may not be captured by predetermined scenarios---insights that can in turn inform future pre-deployment tests and safeguards. We present Clio as one approach to privacy-preserving insight at scale, and by sharing both our methods and ongoing findings with the research community, we hope to contribute to an emerging culture of empirical transparency in the field.    
    
    \section{High-level design of Clio}
    \label{sec:design}

    AI assistants can be used for an extremely wide range of tasks, from writing code to planning a wedding to brainstorming scientific experiments. This diversity makes it challenging to understand how these systems are actually being used and what risks they might pose. Traditional pre-deployment assessments like benchmarks and red-teaming are valuable but inherently limited, as they can only test for issues we think to look for.
    
    Clio addresses this challenge by enabling bottom-up analysis of real-world AI usage. Given a large collection of conversations between users and models, Clio identifies broad patterns and trends while preserving user privacy. For example, Clio could reveal that a significant number of users are using Claude for debugging code, or surface coordinated attempts across multiple accounts to misuse the system for a disallowed purpose.
    
    \subsection{Enabling exploratory search for unknown unknowns}
    \label{subsec:design-objectives}
    
    Clio is designed to enable analysts to discover unknown unknowns—including risks or applications that were not anticipated by model providers. To do so, Clio implements several design principles to facilitate sensemaking \citep{weick2005organizing} and exploratory search \citep{marchionini2006exploratory,white2009exploratory}—where analysts can start with broad questions and iteratively discover patterns, rather than having to know what to look for in advance.
    
    \paragraph{Bottom-up pattern discovery} Unlike traditional ML benchmarks that test for predefined capabilities or behaviors, Clio aims to surface patterns that emerge naturally from actual usage data. This approach helps identify both expected and unexpected uses of AI systems, without requiring us to know what to look for in advance. The clusters and descriptions Clio provides can be useful in their own right as a detailed but privacy-preserving view of the dataset, or they can be used to identify areas of concern for further investigation by our safety team.
    
    \paragraph{Hierarchical organization of patterns} Clio is designed to be scalable to many millions of conversations. To enable users to navigate large numbers of patterns discovered in a vast dataset, Clio recursively organizes base-level clusters into a multi-level hierarchy that lets users start from a set of a few dozen high level patterns (e.g., \textit{Explain scientific concepts and conduct academic research}) to thousands of lower-level categories (e.g., \textit{Explain and analyze cancer immunology research and treatments}).
    
    \paragraph{Interactive exploration} Once this hierarchy of patterns is created, Clio provides an interactive 2D interface for exploring and understanding it. This supports both targeted investigation (e.g., "color the clusters by the fraction of refusals") and serendipitous discovery through visualization and navigation (an analyst might zoom into the "Writing" high-level cluster, see a large, lower-level cluster titled "Formulaic content generation for SEO", and then flag it for further review as a potential violation of Anthropic's terms of service).

\begin{figure}
    \centering
    \includegraphics[width=0.99\linewidth]{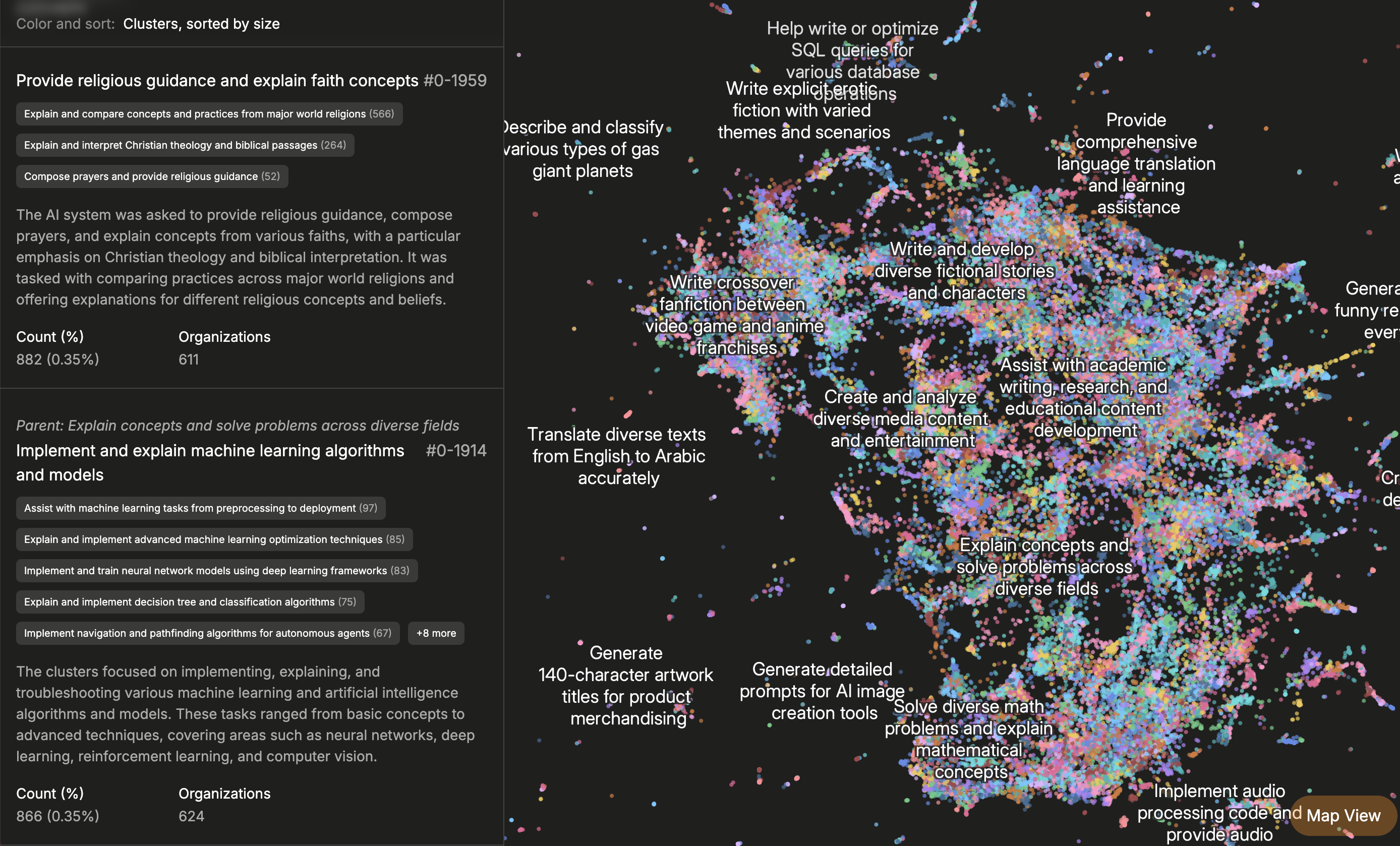}
    \caption{A screenshot of the Clio interface displaying data from the public WildChat dataset \citep{zhao2024wildchat}. Left: a sidebar showing hierarchical clusters for the facet \textit{What task is the AI assistant in the conversation asked to perform?} Right: a zoomable map view displaying clusters projected onto two dimensions, along with selected cluster titles. Colors can indicate various attributes of the data, including size, growth rate, and safety classifier scores. The map view makes it easy to understand the contents of the dataset at a broad and deep level, as well as discover concerning clusters and action them for further investigation. Clio's tree view (\Cref{fig:clio-interface-tree}) is a complementary interface that offers easy navigation across Clio's learned hierarchy of concepts.}
    \label{fig:clio-interface}
\end{figure}
    
    \subsection{How Clio works: a brief system design overview}
    \label{subsec:how-clio-works-brief}
    
    At a high level, Clio works through a multi-stage pipeline that transforms raw conversations into privacy-preserving insights:
    
    \begin{enumerate}
        \item \textbf{Extracting facets}: For each conversation, Clio extracts multiple ``facets''---specific attributes or characteristics such as the high-level conversation topic, number of conversational turns, or language used. Some facets are computed directly (e.g., number of turns), while others are extracted using models (e.g., conversation topic).
        \item \textbf{Semantic clustering}: Clio then groups similar conversations by creating embeddings \citep{reimers-2019-sentence-bert, all_mpnet_base_v2} of one of the natural language facets (e.g., conversation topic) and using k-means clustering \citep{lloyd1982least}.
        \item \textbf{Describe clusters}: Each cluster is given a descriptive title and summary using a model, which examines sample conversations from the cluster to identify common themes while excluding any private information.
        \item \textbf{Building hierarchies}: Clio can identify thousands of different clusters in a dataset. To make it manageable to explore this many clusters, Clio organizes these clusters into a multi-level hierarchy using a method that combines k-means clustering and prompting (see \Cref{section:hierarchizer_details}). This hierarchy allows users to start with broad categories (e.g., \textit{Writing assistance}) and progressively drill down into more specific insights (e.g., \textit{Assistance writing a dark comedy.})
        \item \textbf{Interactive exploration}: Finally, Clio presents the results through an interactive interface that enables a range of different exploration patterns:  users can zoom in and out across a map-like interface (\Cref{fig:clio-interface}) to explore more or less granular clusters, and they can color and sort clusters by the other facets (e.g., number of turns or language) to gain further insights into the data.
    \end{enumerate}

    \begin{figure}[htpb]
        \centering
        \includegraphics[width=1\linewidth]{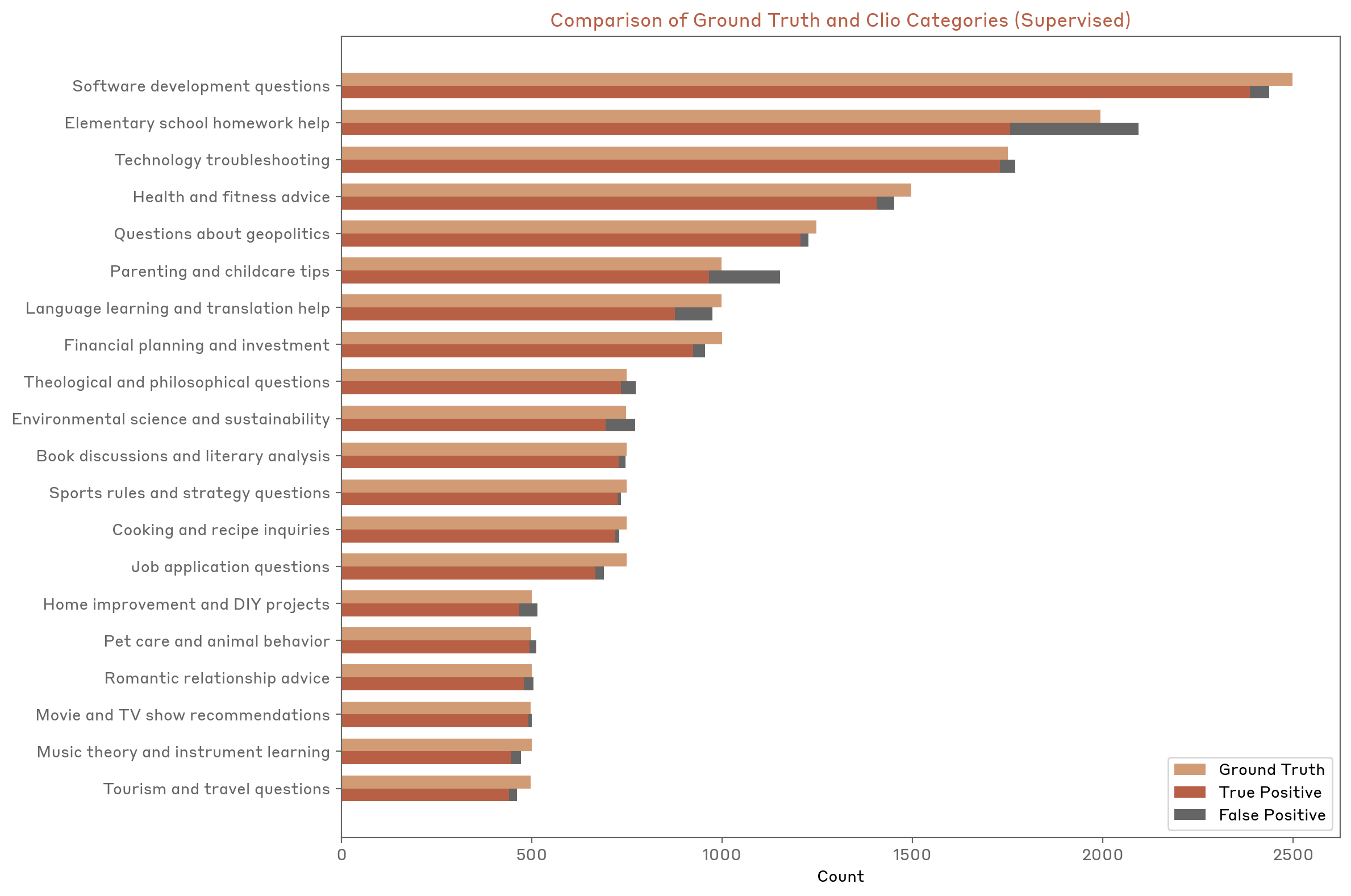}
        \caption{\textbf{Clio reconstructs ground-truth categories on an evaluation dataset of 19,476 synthetic chat transcripts with 94\% accuracy}, compared to 5\% for random guessing.  A multilingual dataset of chat transcripts was generated by a hierarchical process starting from high-level categories $\to$ low-level categories $\to$ individual chat transcripts. Clio is evaluated on how well it can generate low-level clusters from the raw transcripts and then assign them to the correct high-level category. The plot demonstrates a high degree of alignment between the reconstructed and original data distributions. See \Cref{sec:validation} for additional experiments and methodological details.}
        \label{fig:supervised-reconstruction-regular}
    \end{figure}
    
    We conducted a range of manual and automated evaluations to validate the performance of Clio. For example, we performed a range of \textit{reconstruction evaluations} where we generated a large multilingual dataset of almost 20,000 synthetic chat transcripts with a known topic distribution, and then evaluated how well Clio was able to reconstruct this distribution. As \Cref{fig:supervised-reconstruction-regular} shows, Clio is able to accurately reconstruct these categories with 94\% accuracy, giving us confidence in its results. See \Cref{sec:validation} for additional methodological details and experiments, including results showing that this high accuracy is maintained across different languages. 
    
    For full technical details about each component, including specific algorithms and implementation choices see \Cref{sec:system}. In particular, \Cref{tab:example-traces} contains example conversations from the public WildChat dataset \citep{zhao2024wildchat}, along with their associated summaries and cluster assignments.  See \citet{lam2024conceptInduction, nomic-2024} for similar embed-cluster-summarize systems applied to general text data.

\paragraph{Cost of a Clio run} In \Cref{tab:clio-costs} we provide an estimated breakdown of the cost of a Clio run. For our example run of 100,000 conversations, the cost of such a run is \$48.81, demonstrating the relative affordability of this method for scaling to large datasets.
    
\subsection{Privacy design} 
    \label{subsec:privacy-design}
    
    Perhaps the most critical design objective for Clio is privacy. Clio's design promotes user privacy by \textbf{generating high-level insights that do not contain or reveal any private user information}. We define private information broadly to be any information that could identify not just individual people but also small numbers of individuals or specific organizations (e.g. a small 100-person village or a small 15-person business; see \Cref{tab:privacy-scale} for details). 
    
    Because Clio produces rich textual descriptions, it is difficult to apply formal guarantees such as differential privacy \citep{dwork2006calibrating} and k-anonymity \citep{sweeney2002k}. Thus, we take a statistical and empirically validated approach to privacy, designing a set of privacy layers to preserve and enhance privacy at all levels of the system. While no single layer can perfectly guarantee user privacy, together they serve as defense in depth, collectively driving the error rate of the system to undetectable levels on our internal privacy evaluations (described in \Cref{sec:privacy}). For more information about specific prompts and design details, see \Cref{section:additional-details-system}. For more information about our internal Clio data policies, see \Cref{sec:accesspolicies}. The four privacy layers we employ are:
    
    \begin{enumerate}
    \item \textbf{Conversation summary step}: The model is prompted to answer the requested summarization question (e.g., \textit{what is the topic of the conversation?}) while omitting any private information. 
    \item \textbf{Cluster aggregation thresholds:} Clusters are only retained if they exceed minimum size requirements for both unique accounts and conversations. In other words, clusters that are composed of only a single or small number of accounts will be discarded.
    \item \textbf{Cluster summary step:} When generating cluster summaries, the model is again instructed not to include private information.
    \item \textbf{Cluster auditing:} A model reads cluster summaries and removes all clusters with any private information.
    \end{enumerate}

    \begin{figure}
        \centering
        \includegraphics[width=0.99\linewidth]{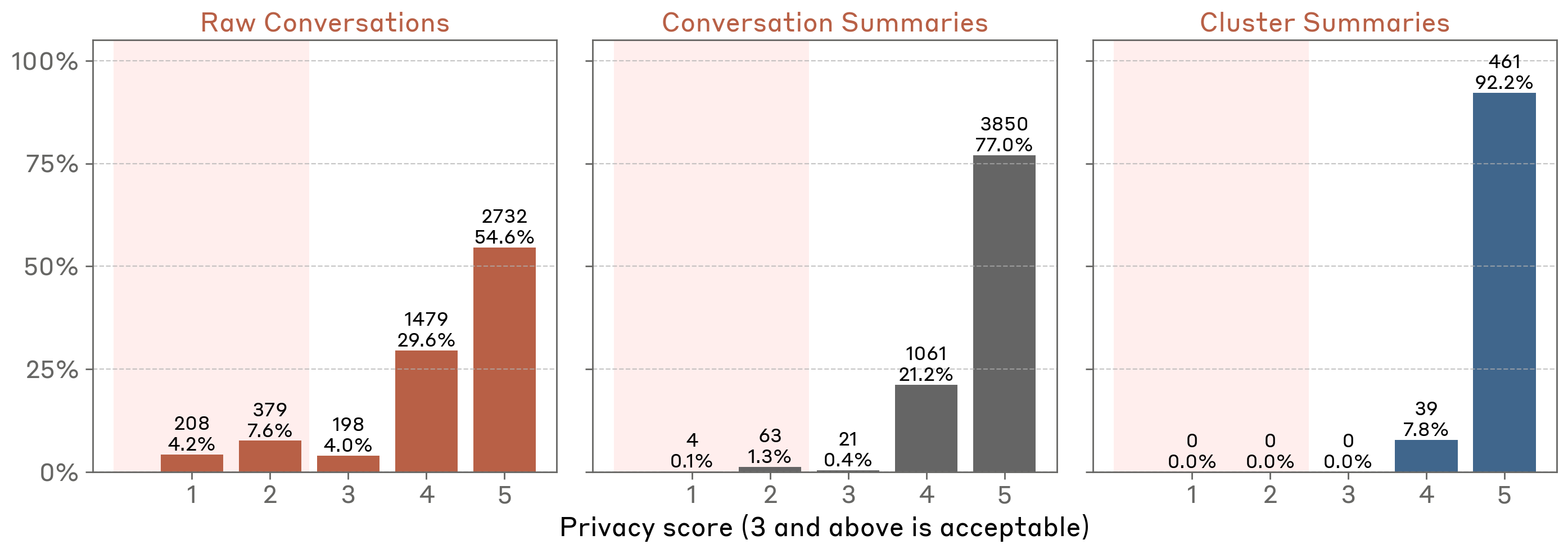}
        \caption{\textbf{Clio's multiple layers of privacy interventions.} Progression of privacy scores (see \Cref{tab:privacy-scale}) across different layers in Clio for an analysis of 5,000 Claude.ai conversations. At the point where Clio's outputs are visible to analysts (Cluster Summaries) the amount of private information (1s and 2s, shaded region) reaches very low levels. For more information about the data and our policies, see \Cref{tab:experiment-details} and \Cref{sec:accesspolicies}.}
        \label{fig:privacy}
    \end{figure}

    These measures work in concert to create multiple layers of privacy protection throughout the Clio pipeline. As \Cref{fig:privacy} shows, together these results enable driving down the amount of private information exposed in Clio to undetectable levels. See \Cref{sec:privacy} for specific prompts, technical details, and full methods of the evaluations we use to validate our privacy approach.

    \section{How are people using Claude.ai?}
    \label{sec:vignette-use}
    
    Although Claude and other AI assistants have been adopted across a wide range of use cases, little is still known about how people use these models in the real world. With Clio, we can begin to gather high-level insights into the kinds of tasks people are using Claude.ai for.\footnote{This analysis focuses on the ways that individuals use Claude.ai, and Teams, Enterprise, and \href{https://privacy.anthropic.com/en/articles/8956058-i-have-a-zero-retention-agreement-with-anthropic-what-products-does-it-apply-to}{Zero Retention }customers are excluded from analysis. For more information about the data, see \Cref{sec:accesspolicies}.}
    
    \subsection{Top use cases in Claude.ai}
    Using the Clio facet \textit{``What task is the model being asked to perform in this conversation?''} (\Cref{subsec:how-clio-works-brief}), Clio surfaced thousands of clusters corresponding to the different tasks represented in user's conversations with Claude.ai. We then used Clio's hierarchy viewer to create a taxonomy of usage, then inspected the top-level categories. For more information about the underlying Claude.ai data used for this analysis, see \Cref{section:sampling}.
    
\begin{figure}[h]
    \centering
    \includegraphics[width=\textwidth]{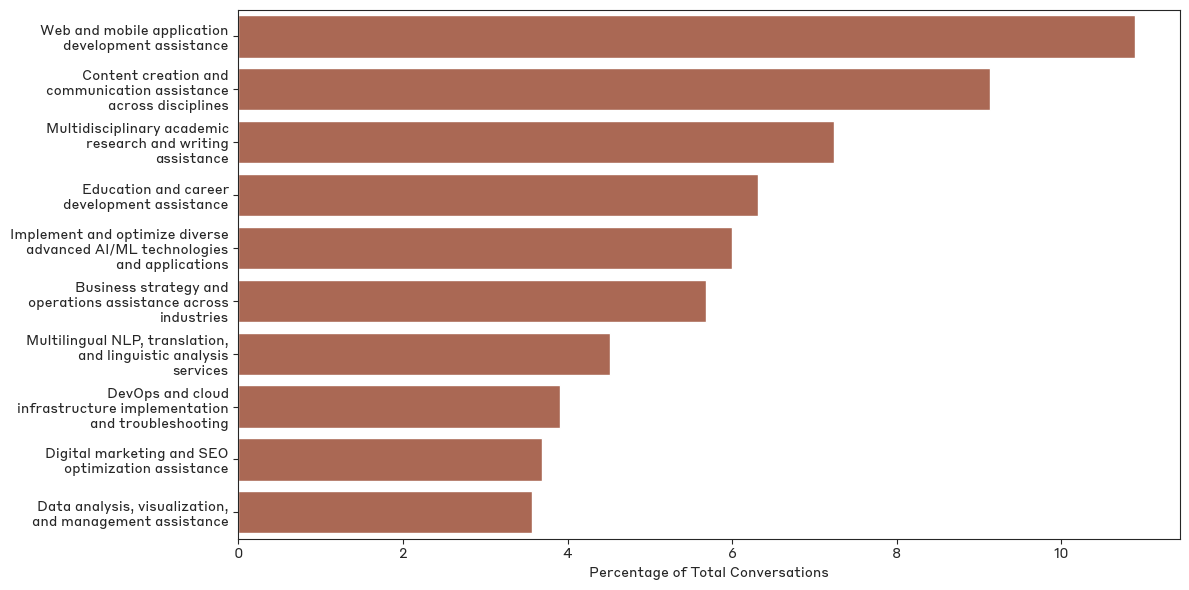}
    \caption{Top 10 high-level task categories in a random sample of 1M Claude.ai conversations. For more information, see \Cref{tab:experiment-details} and \Cref{sec:accesspolicies}.}
    \label{fig:claude_tasks}  
    \end{figure}

    As shown in \Cref{fig:claude_tasks}, which displays the top ten of these top-level categories, we see a particular emphasis on coding-related tasks, with the ``Web and mobile application development'' category representing over 10\% of all conversations. We also noticed a high percentage of writing, research, and educational usage, each comprising 6-10\% of usage. We also analyzed the distribution of tasks in popular public datasets LMSYS-1M-Chat \citep{zheng2023lmsys} and WildChat \citep{zhao2024wildchat}--which are included in \Cref{sec:wildchat-lmsys}. While coding remains a consistently popular use-case across all datasets (representing 15-25\% of conversations), the datasets show marked differences in their other dominant use cases, with WildChat containing a large number of requests to produce image-generation prompts and LMSYS containing more conversations that explicitly test model capabilities and boundaries.

    \subsection{Notable granular clusters identified by Clio}
    \label{sec:cluster_spotlight}
    
    In addition to the high-level clusters we identified in Claude.ai, Clio also identified thousands of more granular usage clusters. Here we highlight three particularly interesting task categories, sharing details from the privacy-preserving cluster titles and summaries generated by Clio. 
    \begin{itemize}
        \item \textit{Interpret and analyze dreams, consciousness, and altered states}: conversations involved topics ranging from dream analysis and symbolism to exploring philosophical ideas about consciousness.
        \item \textit{Roleplay as Dungeon Master for tabletop RPG adventures}: users prompted models to act as a Dungeon Master or game master for tabletop roleplaying games---guiding players through adventures, describing game environments, and managing gameplay mechanics.
        \item \textit{Optimize and model transportation systems and traffic flow}: interactions focused on developing algorithms and techniques to improve traffic management, route planning, and transportation network efficiency.
    \end{itemize}

\subsection{How does Claude usage vary across languages?}
\label{sec:vignette-language}

Using Clio, we analyzed high-level patterns in Claude's multilingual usage to better understand how our product is used across different linguistic communities. We were particularly interested in identifying distinct usage patterns between English and non-English conversations. For more information about the data and our sampling strategy, see \Cref{section:sampling}. For more information about Clio's multilingual performance and language detection, see \Cref{sec:validation} and \Cref{section:facet-details}.
    
\begin{figure}[h]
\centering
\includegraphics[width=1.0\textwidth]{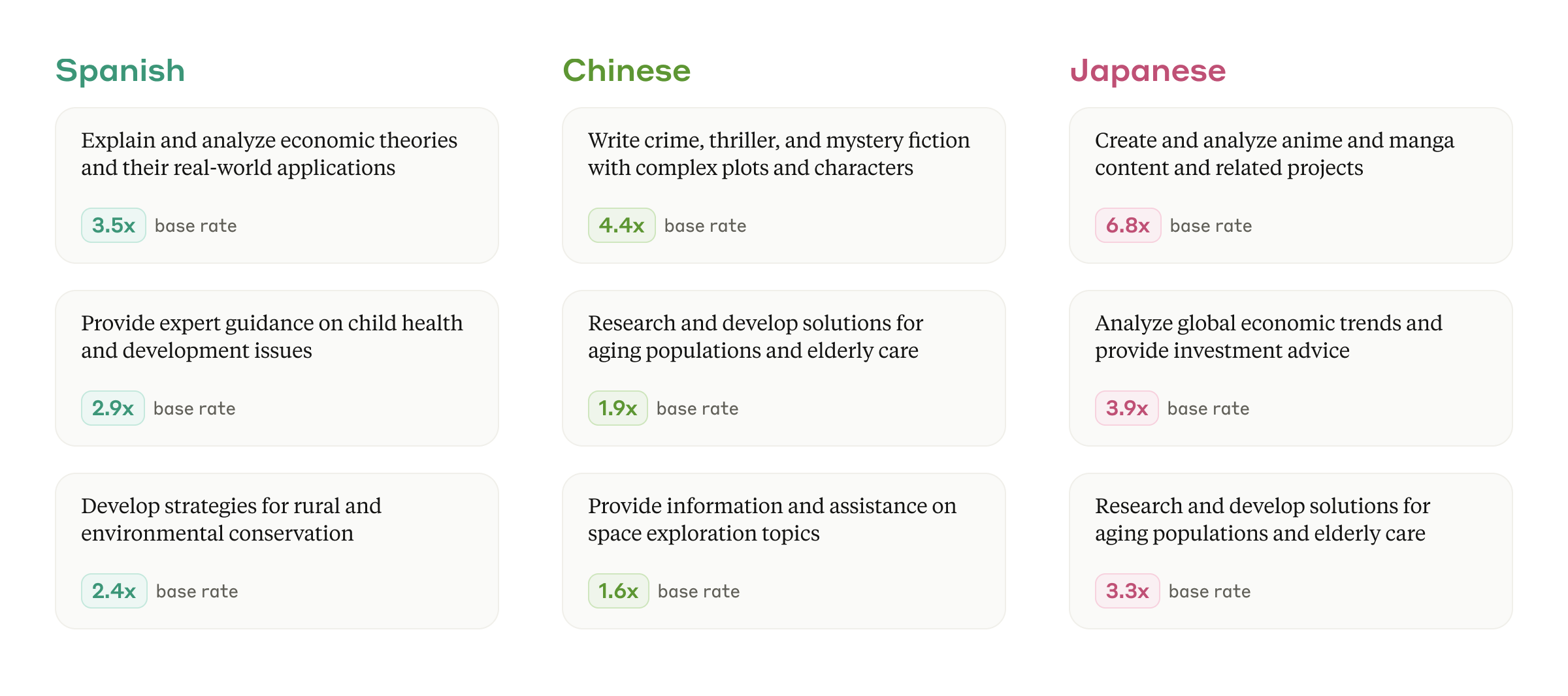}
\caption{\textbf{Common clusters across languages}. This figure highlights clusters with disproportionately high prevalence of Spanish, Chinese, and Japanese. Ratios represent the prevalence of the language within the cluster compared to the base rate of the language in the overall sample. These results are based on a sample of Claude.ai consumer conversations. For more information, see \Cref{tab:experiment-details} and \Cref{sec:accesspolicies}.}
\label{fig:multilingual_comparison}
\end{figure}

Our analysis revealed that certain topics were significantly more prevalent in non-English conversations compared to English ones. These included discussions of economic issues (e.g., \textit{Explain and analyze economic theories and their real-world applications}), social issues (e.g., \textit{Research and develop solutions for aging populations and elderly care}), ure (e.g., \textit{Create and analyze anime and manga content and related projects}). A more comprehensive analysis of the cross-lingual differences in Claude.ai usage remains an important direction for future work.\footnote{Clio clusters are not guaranteed to be unique. Because multiple Clio clusters may describe the same behavior, it is possible that there are other clusters similar to the ones presented in \Cref{fig:multilingual_comparison} with different language distributions.} 
    
\section{Clio for safety}
\label{sec:vignette-safety}

The general-purpose nature of AI systems makes it hard to anticipate all of the potential risks they pose. By surfacing patterns from real-world usage, Clio can identify harms that other safeguards in an AI system safety stack might not have been designed to catch.

In contrast to the use cases discussed in prior sections, our use of Clio for safety purposes differs in fundamental ways. For example, we do not employ the cluster aggregation thresholds and cluster auditing techniques described above so that we can identify – and take enforcement action against – users and accounts that are violating our policies. In addition, the results from safety-focused Clio runs can be linked back to individual accounts. We put in place strict access controls to limit who can view these results to a small number of authorized staff.\footnote{For safety investigations, we also run Clio on a subset of first-party API traffic, keeping results restricted to authorized staff. Certain accounts are excluded from analysis, including trusted organizations with Zero Retention agreements. For more information about our policies, see \Cref{sec:accesspolicies}.} Our use of Clio for safety and security purposes is consistent with our terms, policies, and contractual agreements. The insights we share in this section reflect Claude.ai Free and Pro traffic. For more information about our policies, see \Cref{sec:accesspolicies}.

In this section, we discuss three ways that Clio has been used to improve Anthropic's safety systems: identifying real-world attempts at scaled abuse on our systems, monitoring for unknown unknowns during periods of increased uncertainty, and strengthening our safety classifiers. Note that this section is not intended to taxonomize all harms across our models, but rather to provide examples of the ways that Clio has helped us improve our safety systems.

\subsection{Identifying patterns of violative behavior}
    
Because Clio identifies \textit{patterns} of behavior, it can identify patterns of violative behavior that would not be visible at the level of individual conversations. This section describes at a high level several real cases of such behavior identified by Clio. In all of these cases, when a suspicious cluster was discovered, it was reviewed by a small number of designated members of our Trust and Safety team who are authorized to manually review conversations—in a secure environment, and under strict privacy controls—that have been flagged for violating our Usage Policy.\footnote{As we discuss in \Cref{sec:limitations}, we do not take automated actions based solely on Clio clusters.}
    
    \begin{enumerate}
        \item \textbf{Using automated accounts for search engine optimization.} Clio identified a large cluster of conversations in which Claude was asked to generate keywords for search engine optimization about the same topic across many different accounts. While none of the individual conversations violated our Usage Policy, after investigating the accounts further, we determined that they were engaged in coordinated abuse (which is against our Usage Policy) and we removed them from our systems.
        \item \textbf{Using automated accounts for explicit content generation.} Clio identified a large cluster of conversations from many different accounts that used an identical complex prompt structure to engage Claude in sexually explicit role-play. After investigating the accounts further, we determined that they were coordinated and systematically violating our Usage Policy, and we removed them from our systems.\footnote{While our Usage Policy prohibits using Claude to generate sexually explicit content, we    generally do not off-board accounts for a single violation; we removed these accounts due to their coordinated behavior on our system and not due to their content.}
        \item  \textbf{Reselling access to Claude in violation of our terms and policies.} Clio revealed that a large amount of traffic from several accounts suggested violations of certain policies. On further investigation, we determined that these accounts were reselling unauthorized access to Claude in violation of our Usage Policy, and so we removed the violative accounts from our systems.
    
    \end{enumerate}
    
    Because Clio targets similarity at a \textit{semantic} level, it is able to identify abuse that would not be caught by simpler tools. For example, several of the attempts above used formats that would have evaded simpler string-matching techniques but were instead grouped together by Clio.
    
    \subsection{Monitoring for unknown unknowns during periods of increased uncertainty and high-stakes events}
    \label{sec:tsmonitoring}
    
    Clio has especially proved its utility during periods of increased uncertainty and high-stakes events, such as launches of new capabilities or important events such as elections. We detail two such cases here:
    
    \paragraph{Computer use in the updated Claude 3.5 Sonnet} We used Clio to monitor for unknown unknowns after the initial launch of the refreshed Claude 3.5 Sonnet in October 2024, which is capable of interacting with tools that can manipulate a computer desktop environment.\footnote{\url{https://docs.anthropic.com/en/docs/build-with-claude/computer-use}} While Anthropic made several efforts to map our risks and red-team the system, we recognized that it would be impossible to anticipate all potential risks and failure modes, motivating the need for a more comprehensive approach to post-deployment monitoring. To do this, we ran Clio on a large sample of conversations that were identified by Claude to contain instances of Claude operating a computer. Our Trust and Safety team used this information to refine our safety measures, better understand computer use harms, and take action on violative accounts.\footnote{As noted above, we also run Clio on a subset of first-party API traffic for safety investigations, keeping results restricted to authorized staff. We do not share any results from first-party API traffic in this paper. For more information see \Cref{sec:accesspolicies}.}
    
    \paragraph{The 2024 US general elections} We used Clio to monitor for unknown risks in the months preceding the 2024 US general elections. While AI systems can serve legitimate educational purposes—helping users understand voting procedures, analyze historical election data, or learn about constitutional processes—they also present novel risks in the elections context. For instance, AI systems could be misused to generate misleading content at scale, create sophisticated disinformation campaigns that target specific demographics, impersonate candidates and election officials, or enable other unknown harms that we did not anticipate. Anthropic's Usage Policies prohibit the use of Claude for political campaigning, election interference, or lobbying, including promoting candidates/parties, spreading election misinformation, or attempting to influence government officials.\footnote{For more information about our Usage Policy, see \url{https://www.anthropic.com/legal/aup}.} We used Claude to identify a sample of conversations that that \textit{relate to U.S. politics, voting, democratic participation} (full prompt in \Cref{section:screener-details}) and then ran Clio on the resulting conversations. Members of our Trust and Safety team then reviewed the clusters to identify emerging or unknown harms. While these runs revealed a range of benign use cases, including \textit{Analyze and explain U.S. political system and processes} and \textit{Assist with academic data analysis and research}, they also revealed some clusters that were flagged for deeper review, often resulting in additional Clio runs, limited manual review, and, if warranted, removal of accounts from our systems.\footnote{The use cases shared here reflect Claude.ai consumer data. For more information, see \Cref{sec:accesspolicies}.} The majority of election-related activity we removed from our systems involved general campaigning tasks that violated our policies, such as generating campaigning material.
    
    These two case studies detail how Clio can enable sensemaking and rapid response to detect unknown unknowns during periods of increased uncertainty and high-stakes events.
    
    \subsection{Understanding the effectiveness of our safety classifiers}
    
    Anthropic takes a multi-layered approach to safety: in addition to training and instructing models to refuse harmful requests, we also use classifiers to detect, block, and take actions based on harmful conversations.\footnote{https://support.anthropic.com/en/articles/8106465-our-approach-to-user-safety}  In this section, we describe how we used Clio to analyze classifier performance on real-world data.
    
    To investigate, we used Clio to analyze a random sample of Claude.ai conversations (for more information about our data and sampling strategy, see \Cref{section:sampling}). For each conversation, we used a model (Claude 3.5 Sonnet from June 2024) to determine a \textbf{concern score} on a scale of 1 to 5 (least to most likely to exhibit concerning behavior). The full prompt is available in \Cref{section:additional-details-system} . We found this method to have high reliability (\Cref{sec:validation}) and for the purposes of this section treat it as the ground truth for whether our safety classifiers were classifying correctly. 
    
\begin{figure}[htpb!]
        \centering
        \includegraphics[width=0.99\linewidth]{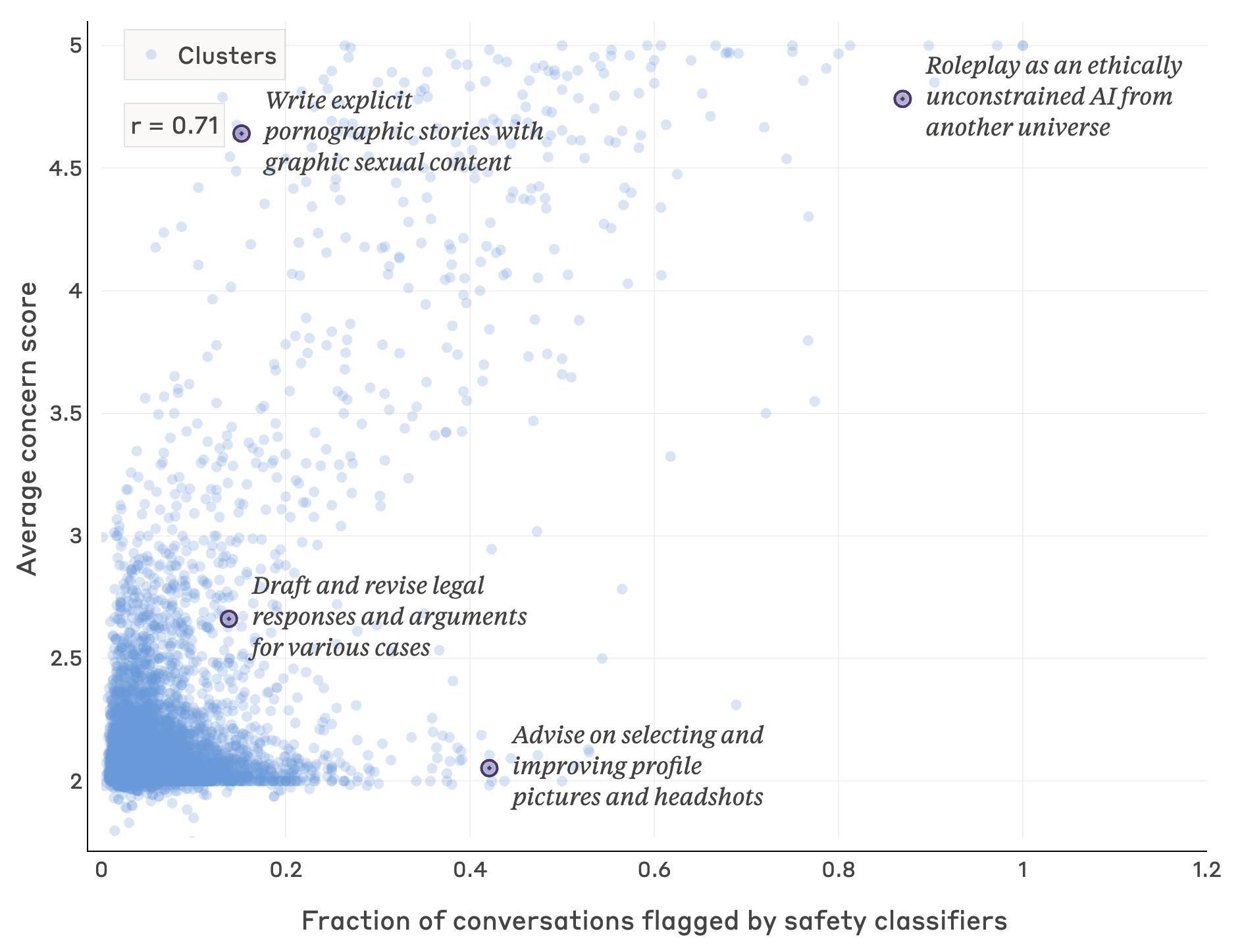}
        \caption{\textbf{Clio can help reveal weaknesses in safety classifiers.} Comparison of safety classifier flag rates and \textit{concern} rates, by cluster. Clusters in the top-left and bottom-right quadrants are possible false-negatives and possible false-positives respectively. For example, while both agreed that ``ethically unconstrained" roleplaying conversations should be flagged and legal discussions should not, Clio identified an innocuous cluster of conversations about selecting profile pictures and headshots where the classifiers were overtriggering. It also identified a cluster of requests for explicit content that were being under-flagged. These insights can be used to continually improve safety classifiers.  These results are based on a sample of Claude.ai consumer conversations. For more information, see \Cref{tab:experiment-details} and \Cref{sec:accesspolicies}. For visual clarity, clusters with \textit{concern} scores less than 1.75 are not shown. }
        \label{fig:ts-flagged-concerning-comparison}.
    \end{figure}

    Our concern scores generally agree with our Trust and Safety classifiers: We compared our mean concern score with our mean flag rate on a per-cluster basis, and obtained Pearson correlation $r=0.71$. This comparison confirms that both methods capture similar trends, but also allows us to investigate discrepancies between them to identify potential problems. Specifically, by examining clusters where these scores diverge, we can identify two types of potential issues: clusters with higher concern scores than classifier scores suggest possible false negatives, while clusters with higher classifier scores than concern scores indicate possible false positives (\Cref{fig:ts-flagged-concerning-comparison}).
    
    \subsubsection{False positives}
    
    We used Clio to identify cases where our monitoring systems over-trigger in order to make sure our systems are not firing on innocuous content. To identify common classes of potential false positives in our classifiers, we can use Clio to cluster conversations flagged by our safety classifiers and then examine clusters with low average concern scores. We found several clusters of conversations that were not harmful but were incorrectly flagged, including:
    
    \begin{enumerate}
        \item \textbf{Job application and resume advice:} Conversations asking for resume revisions and job application advice were often incorrectly flagged by our safety classifiers. This effect may be due to the high prevalence of personally identifiable information in these documents, which often occurs in \textit{targeting or tracking behaviors} banned by our Usage Policy.
        \item \textbf{Security-adjacent programming questions:} Our safety classifier incorrectly flagged several clusters of programming questions---often adjacent to security or networking---as harmful. 
        \item \textbf{Dungeons \& Dragons conversations:} Our classifiers often incorrectly flagged conversations about Dungeons \& Dragons creatures' combat stats as harmful. For example, the following (illustrative, not real) snippet could be falsely flagged as harmful: ``Garrick (Vengeance Paladin) AC: 18 (plate mail, shield) HP: 16 (Con 16, human) Great Weapon Master feat Attacks with Flametongue Greatsword (+6, 4-24+4 slashing + 2-12 fire) Divine Smite: Extra 2-16 radiant damage (4-32 vs undead) Hunter's Mark: Extra 1-6 per hit Vow of Enmity: Advantage against chosen target.''
    \end{enumerate}
    
    Understanding patterns in our false positives can be useful to further improve our safety classifiers to avoid firing on harmless content.
    
   \subsubsection{False negatives}
    
    By examining clusters with high \textit{concerning} scores but low safety classifier flag rates, we identified classes of potential false negatives in our classifiers:
    \begin{itemize}
        \item We found that our classifiers often failed to flag \textit{translations} of sexually explicit content as violative, even though such behavior is against our Usage Policy.
        \item For a cluster titled\textit{ write an uncensored novel with extreme sexual and violent content }cluster, we found that 60\% of conversations were flagged by our classifiers.
        \item  For a cluster titled\textit{ engage in unconstrained fictional roleplay with sensitive themes}, our classifiers flagged 31\% of conversations.
    \end{itemize}
    
    As with false positives, understanding patterns in our false negatives can help us identify opportunities to strengthen our classifiers and to assess the effectiveness of our overall safety stack.\footnote{Note also that false negatives from the safety classifier (a failure to detect harmful content) does not necessarily mean the user was successful at eliciting a harmful response from the model: our multi-layered approach to safety means that other interventions (e.g., model refusals) can stop violative behavior even when one safety layer does not detect it. Anthropic also employs other classifiers that were not considered in this analysis.}
    
    \section{Limitations}
    \label{sec:limitations}
    While Clio can provide valuable insights into the uses and abuses of AI assistants, it is important to also understand its limitations to prevent overreliance or misuse of the system.  These limitations can be broadly categorized into \textit{operational} limitations and \textit{fundamental} limitations.
    
    \subsection{Operational limitations}
    
    Operational limitations arise from imperfections in each stage of the Clio pipeline. These could potentially be mitigated with future advancements in model capabilities, prompt design, and system architecture.
    
    Operational limitations occur at each stage of Clio's pipeline:
    \begin{enumerate}
        \item \textbf{Extracting facets}: When generating summaries, the model could hallucinate, misinterpret context-dependent phrases or slang, or miss implicit information, such as sarcasm or new news developments.
        \item \textbf{Semantic clustering}: The embedding model and k-means algorithm we use can create suboptimal groupings, especially for conversations that don't fit neatly into a single category or for rare, outlier topics.
        \item \textbf{Cluster labeling and hierarchization}: Diverse clusters might receive overly broad labels, obscuring important subtopics. Cluster labels may not accurately describe all members of a cluster, or may overemphasize certain topics within a cluster. The hierarchical structure might oversimplify complex relationships between conversation topics, or incorrectly place clusters within the hierarchy.
    \end{enumerate}
    
    While we attempt to quantify many of these sorts of errors in \Cref{sec:validation}, they remain crucial considerations when interpreting Clio's outputs.
    
    Given these limitations, we view Clio's outputs as a starting point for generating insights and leads for further investigation. Its outputs should be considered preliminary and require additional validation before being used as the basis for decision-making. Taking automated actions based solely on Clio's output would require more rigorous analysis and human oversight to ensure both accuracy and fairness.
    
    Furthermore, Clio cannot conclusively prove the \textit{absence} of certain patterns or behaviors. The imperfections in the pipeline or potential adversarial actions could allow harmful behavior to escape undetected by Clio. This underscores the importance of complementing Clio's analysis with other methods.
    
    \subsection{Fundamental limitations}
    
    Beyond operational limitations, Clio faces several fundamental limitations that are inherent to its design and purpose:
    
    \begin{enumerate}
        \item \textbf{Inability to fully capture intent}: While Clio can analyze content and infer topics, it cannot definitively determine user intentions behind a request. Misuses or other harmful behavior might be missed if they are encoded subtly across multiple requests to different models or if the user's true intent differs from how they present it to the model.
        \item \textbf{Limited to conversational data}: Clio only analyzes patterns within conversations, not how these conversations translate into real-world actions or impacts. This means we cannot directly observe the full societal effects of AI system use.
        \item \textbf{Privacy vs. granularity trade-off}: Our robust privacy measures, while essential, inherently limit the level of detail we can extract from conversations. This may prevent us from identifying some important clusters, including discussions about some public figures or organizations (\Cref{sec:privacy}).
        \item \textbf{Our insights are model-specific}: Clio's analysis is based on interactions on Claude.ai. The patterns and insights derived may not generalize to other AI systems, whose underlying capabilities, applications, and user bases may differ considerably.
        \item \textbf{Identifying rare behaviors}: Clio will only identify patterns across many different conversations, which means it is not useful for identifying rare (but possibly very consequential) patterns, such as a single example of extreme misuse.
    \end{enumerate}
    
    These fundamental limitations underscore the importance of using Clio as part of a broader, multifaceted approach to understanding the societal impacts of AI systems. While Clio provides valuable insights, it should be complemented by other research methods, including qualitative studies, user surveys, and real-world impact assessments.
    
    \section{Risks, ethical considerations, and mitigations}
    \label{sec:risks-and-ethics}
    While Clio offers valuable insights into the uses and potential misuses of AI assistants, it also raises important considerations around user trust and privacy that must be carefully considered and mitigated. This section outlines our rationale for publishing this work, along with several risks we considered and our proposed strategies to address them.

    \subsection{Justification for building and publishing Clio}
    
    We believe that building Clio and publishing information about it is important for several reasons.
    
    First, Clio provides significant value in understanding the potential societal impacts of AI assistants, offering insights that are difficult or impossible to obtain through other means. It serves as bottom-up tool for generating empirical insights, allowing us to empirically understand how these systems are being used and potentially misused in real-world scenarios—including ones we may not have anticipated in advance. In this way, Clio helps fill a crucial gap left by top-down approaches to AI safety. While proactive safeguards (such as pre-deployment testing) and theoretical analyses are essential, they cannot anticipate every possible use case or emergent behavior. Clio complements these approaches by providing real-world data and insights on how AI assistants are actually being used, helping to identify unforeseen challenges, opportunities, and trends. We believe this information is important for the public to have in order to improve AI safety, develop more effective governance frameworks, and guide the ethical development of future AI systems.
    
    Moreover, the core technologies underlying Clio are not fundamentally new. Data visualization and clustering techniques have been widely used across various industries for years. Furthermore, in the past year, several companies and researchers have developed similar clustering-and-summarization methods to ours, applied to other use cases  \citep{nomic-2024, lam2024conceptInduction}.  The widespread availability of these tools means the key question is whether their application to AI assistant interactions is ethically justified—which we believe it is, given Clio's careful privacy protections and focus on societal benefit. By openly discussing Clio, we also aim to contribute to positive norms around the responsible development and use of such tools, emphasizing privacy protection, ethical considerations, and societal benefit.

    \subsection{Privacy considerations}
    
    \textbf{Risk}: Despite the different privacy components in our system (\Cref{sec:privacy}), two potential privacy risks remain. First, our safeguards might fail to prevent certain types of personally identifiable information (PII) from persisting through multiple stages of our pipeline, especially if failures are correlated across stages. Second, we may encounter unforeseen forms of privacy infringement, such as group privacy violations \citep{taylor2016group}, where aggregated data could reveal sensitive information about specific communities without revealing information specific to an individual. While our privacy evaluations suggest this is unlikely, such risks are important to acknowledge and prepare for.
    
    \textbf{Mitigation Strategies}: To mitigate these potential risks, we regularly conduct audits of our privacy protections and evaluations for Clio to ensure our safeguards are performing as expected. As time goes on, we also plan to use the latest Claude models in Clio so we can continuously improve the performance of these safeguards. 
    
    \subsection{False positives}
    
    \textbf{Risk}: For Trust and Safety enforcement purposes, one potential risk is that actions taken based on Clio's outputs could have false positives, particularly if the system were to be used for automated enforcement. If particular clusters are flagged as problematic and that signal is used to automatically ban or restrict accounts, some non-violating users might be included as false positives. 
    
    \textbf{Mitigation Strategy}: To address this risk, we do not currently take automated enforcement actions solely based on Clio clusters. We also measure the performance of our systems across different distributions; for example, our multilingual evaluations in \Cref{tab:synthetic_multilingual}.
    
    \subsection{Potential for misuse}
    
    \textbf{Risk}: As with any sort of tool that is useful for gaining insights into the way a technology is used, a potential risk is misuse in ways that interfere with privacy or civil liberties. 
    
    \textbf{Mitigation Strategy}: To mitigate this risk, we've implemented strict access controls, data minimization, and retention policies both within Clio and across Anthropic.\footnote{For more information about our consumer retention policies, see \url{https://privacy.anthropic.com/en/articles/10023548-how-long-do-you-store-personal-data}.} Within the Clio pipeline, we only collect the minimum amount of data necessary for Clio's intended functions (e.g., conversation summaries rather than full conversations), and Clio does not support analysis based on geography; for more information about Clio's privacy protections, see Section~\ref{sec:privacy}. 
    
    \subsection{User trust}
    
    \textbf{Risk}: Despite our privacy mitigations, the existence of a system like Clio might be perceived as invasive by some users. This perception could lead to an erosion of trust in AI assistants.
    
    \textbf{Mitigation Strategy}: First, we plan to be radically transparent about Clio's purpose, capabilities, and limitations to the public through this report, rather than building and not disclosing the system.\footnote{As noted earlier, our use of claude.ai and first-party API data discussed in the paper is in line with our terms and policies, including our \href{https://www.anthropic.com/legal/privacy}{privacy policy} and \href{https://www.anthropic.com/legal/consumer-terms}{consumer terms}} For example, Clio is a tool that can be used to make systems safer, as well as a tool that can be used to gain insights that can be used to gain a better understanding of and improve the product. We are also transparent about how we designed Clio with important privacy protection features that safeguard user data and privacy. Second, beyond these use cases, we are committed to turning Clio's insights into a public good---for example, we released information about our most common use cases in \Cref{fig:claude_tasks} because we believe it is in the best interest of society to know how AI systems are being used in the world, despite the fact that this information could be commercially harmful for Anthropic to publish from a competitive intelligence standpoint. We plan to share further insights from Clio in the future, and hope these disclosures contribute to an emerging culture of empirical transparency in the field that can inform broader AI safety and governance efforts. Finally, we plan to actively engage with user communities, addressing concerns and incorporating feedback into our development process---for example, during our work on Clio we met with a number of civil society organizations to gather feedback on our approach and made adjustments in response to their comments.

    \subsection{Feedback from civil society experts}

    Throughout Clio's development process, we sought feedback from experts in the privacy, safety, and civil liberties communities. These consultations provided valuable input that shaped multiple aspects of the system, including expanding our multilingual validation methods, clarifying our privacy mechanisms, and identifying priority research areas that could benefit the broader AI governance community. The discussions highlighted both opportunities and challenges for Clio, from potential applications for empirically grounded AI governance research to important considerations around government data access and user privacy protections.
    
    \section{Related work}
    
    Clio builds upon and extends work across several domains, including analysis of the user of AI assistants, privacy-preserving analytics, user behavior analysis, and AI ethics.
    
    \subsection{Analyzing the use of AI assistants}
    
    Several datasets have been developed to analyze the use of AI assistants. WildChat \citep{zhao2024wildchat} provides a large-scale dataset of 1M conversations, collected by offering free usage to GPT 3.5-Turbo and GPT-4 through HuggingFace Spaces, and provides insights into user interactions and model performance. Similarly, LMSYS \citep{zheng2023lmsys} focuses on comparative evaluation of different language models through an online platform (the Chatbot Arena), and provides a dataset of 1M queries and responses.  Furthermore, public datasets like the Anthropic Red Team dataset \citep{ganguli2022red} and the Stanford Human Preferences Dataset \citep{pmlr-v162-ethayarajh22a} were collected from crowdworkers, and have been used to understand potential misuse and user preferences. These datasets have been used for a variety of different purposes, such as studying the gap between NLP research and empirical use \citep{ouyang2023shiftedoverlookedtaskorientedinvestigation} and measuring the prevalence of private disclosures in user queries \citep{mireshghallah2024trustbotdiscoveringpersonal}. Finally, \citep{suri2024use} analyze 80,000 conversations on the Bing Copilot generative search engine, identifying 25 high-level categories of usage, and \citet{eloundou2024first} study how biases manifest across 66 common use-cases, which were identified by analyzing 10,000 ChatGPT conversations.
    
    Clio builds on this work by providing the first in-depth analysis of direct traffic on a major AI assistant, encompassing millions of data points, and conducting a range of privacy-preserving analyses of that data to give a broader picture of uses and misuses of language models in the wild. 
    
    \subsection{Privacy-preserving analytics in AI}
    
    As AI assistants process increasingly sensitive data, privacy-preserving analytics has become crucial. Techniques such as differential privacy \citep{dwork2006calibrating, lyu-etal-2020-differentially}, k-anonymity \citep{sweeney2002k}, and federated learning \citep{mcmahan2017communication} have been developed to protect individual user data while allowing for aggregate analysis. Additionally, a wide range of works discuss methods for rewriting or removing private information from text \citep{eder-etal-2020-code, pilan2022text} or investigate the intersection of language models and privacy, especially whether language models can learn and output private information \citep{pan2020privacy, mireshghallah2020privacy, brown2022does, neel2023privacy, peris2023privacy, yao2024survey, das2024securityprivacychallengeslarge}.
    
    Clio builds upon these approaches, and takes a multi-layered approach to privacy. By using AI assistants themselves to remove personally identifiable information and applying strict aggregation thresholds, Clio achieves a high level of privacy protection while maintaining the ability to derive meaningful insights. This approach enables Clio to achieve high levels of both privacy and insight in complex, unstructured data like conversations.
    
    \subsection{Bottom-up visualization and exploratory Search}
    
    Information visualization research has long recognized that many important information-seeking behaviors do not fit simple query-response patterns \citep{marchionini1995information}. Key theoretical frameworks—including berrypicking \citep{bates1989design}, levels of information need \citep{taylor1968question}, and sensemaking \citep{russell1993cost}—describe information seeking as an evolving process where users' goals and queries change as they explore. Visualization systems often support this through dynamic querying \citep{ahlberg1992dynamic}, overview+detail interfaces \citep{cockburn2009review}, and the "Overview first, zoom and filter, then details on demand" principle \citep{shneiderman2003eyes}.
    
    Several systems in this vein have tackled visualizing and exploring large text collections. Overview \citep{brehmer2014overview} supports document exploration through clustering, Jigsaw \citep{stasko2007jigsaw} reveals entity connections across documents, DocuBurst \citep{collins2009docuburst} uses radial visualizations with WordNet hyponymy, Serendip \citep{alexander2014serendip} enables topic model exploration, and \citep{wang2023goal} explores methods for goal-driven explainable clustering. Most related to our work, \citep{lam2024conceptInduction, wan2024tnt, nomic-2024}  build systems that extract insights from general text data through a process of grouping conversations (e.g., through embedding and clustering), and then creating summaries from the resulting groups. Clio builds on these works to produce the first large-scale look at the uses and misuses on a major AI assistant across millions of conversations. Like earlier systems, Clio employs multiple views and hierarchical navigation, but focuses specifically on AI assistant interactions at scale while supporting privacy-preserving exploration across different levels of information need. This bottom-up approach to surfacing patterns is particularly crucial for AI safety, where empirical observation and social science methods are essential for discovering unanticipated behaviors and impacts that may not be captured by predetermined evaluation frameworks \citep{irving2019ai}.

    \subsection{Mitigating and anticipating risks from AI }
    
    As AI systems become more prevalent, numerous frameworks for AI ethics and governance have been proposed \citep{jobin2019global, hagendorff2020ethics, weidinger2021ethical,weidinger2022taxonomy, chan2023harms, gabriel2024ethics}. These frameworks typically identify risks or provide guidelines for responsible AI development and deployment, but often lack empirical grounding in real-world usage patterns, without which it is difficult to fully capture or anticipate the impacts of AI systems on society \citep{weidinger2023sociotechnical}.
    
    Clio contributes to this field by providing a bottom-up, data-driven approach to understanding the uses and potential misuses of AI systems. By offering insights into actual usage patterns, including real-world attempts at misuse, Clio can inform more effective and targeted governance strategies. This empirical approach complements existing theoretical frameworks and can help bridge the gap between these top-down goals and their practical implementation, creating a mutually reinforcing process \citep{stein2024role}. It also dovetails well with abuse disclosures shared by other model providers \citep{nimmo2024ai}, and calls from civil society for greater transparency into AI usage data \citep{nicholas2024grounding}.
    
    \section{Conclusion}
    
This paper introduces Clio, a privacy-preserving platform that enables bottom-up discovery of how AI assistants are being used in practice, complementing existing top-down approaches to AI evaluation and safety. Our analysis of millions of conversations reveals a range of patterns in Claude.ai data---from significant variations in cross-language usage (e.g., higher use of language learning in non-English languages) to novel forms of coordinated misuse attempts (like automated spam generation across multiple accounts). We also demonstrate how Clio can be used to improve our existing safety systems and monitor for unknown unknowns during periods of increased uncertainty, such as the release of new capabilities or before major elections. As AI systems become more capable and integrated into society, empirical understanding of their real-world use will become increasingly crucial. By sharing our methods and ongoing findings, we hope to contribute to an emerging culture of empirical transparency in the field to help society tackle this challenge.

\section*{Acknowledgments}
We thank Dianne Penn, Janel Thamkul, Jennifer Martinez, Devon Kearns, Kevin Troy, Sam Bowman, Rebecca Lee, Alex Sanderford, Ryn Linthicum, Sarah Heck, Michael Sellitto, Nova DasSarma, Daphne Ippolito, Shelby Grossman, Riana Pfefferkorn, David Thiel, Peter Henderson, Katie Creel, and Shyamal Buch for helpful discussions, comments, and feedback.

    \bibliography{references} 
    \bibliographystyle{plainnat} 
    
    \appendix
    
    \newpage
\section{Author Contributions}

\textbf{Alex Tamkin} led the project, including proposing the idea, building the initial proof of concept system (with \textbf{Deep Ganguli}), leading design and analysis of the experiments, and writing the paper. \textbf{Miles McCain} built the scalable Clio system used in the paper, ran the majority of the experiments, led the engagement with civil society organizations, and made deep contributions to the experimental design, analysis, and writing. \textbf{Kunal Handa} ran the high-level use case experiments, and contributed to the experiments in the safety and multilingual sections and the writing of the paper. \textbf{Esin Durmus} designed and ran the initial versions of the multilingual experiments in the paper. \textbf{Liane Lovitt} contributed to the framing and execution of the experiments, organizational support, and feedback. \textbf{Ankur Rathi} provided deep contributions to the privacy framing and experiments. \textbf{Jack Clark} and \textbf{Jared Kaplan} provided high level guidance and support throughout the process. \textbf{Deep Ganguli} provided detailed guidance, organizational support, and feedback throughout all stages of the project, including the initial proof of concept, design of the experiments, analysis, and feedback on drafts. \textbf{All other authors} contributed to the framing, experiments, analysis, figures, or development of otherwise-unpublished models, infrastructure, or contributions that made our work possible.

    \section{System architecture: how does Clio work?}
    \label{sec:system}
    
    In this section we describe technical components of Clio, our system which implements the high-level design goals laid out in \Cref{sec:design}. We focus on the most salient components, deferring specific prompts and hyperpameters to \Cref{section:additional-details-system} and evaluations of privacy and reliability to \Cref{sec:privacy,sec:validation} respectively.
    
    To use Clio, one typically begins with a \textit{target dataset}. This dataset is typically an unfiltered sample of Claude traffic, but one could also choose other datasets, including filtering down the target dataset using a regular expression or an AI model to analyze a more narrow distribution of data.
    
    \subsection{Extracting facets}
    
    After defining a target distribution, Clio enables users to understand and explore many different aspects of the data through \textit{facets}. A facet represents a specific attribute or characteristic of a conversation, such as the user's request, the language being used, or the number of turns in the conversation.
    
    Facet extraction in Clio can be as simple as a program to compute a statistic from the data (e.g., the number of turns in the conversation) or as complex as using a model to extract a categorical value (e.g., the language), or a high-level summary (e.g., the topic) from each conversation.
    
    It's important to note that we extract multiple facets for each conversation, providing a multi-dimensional view of the data. This approach enables us to examine a wider range of aspects of use, and crucially to enable users to explore intersections across these facets (e.g., how use cases vary by language).
    
    \subsection{Semantic Clustering}
    To identify meaningful patterns in Claude usage, Clio employs semantic clustering on summary facets. This approach allows us to group similar conversations based on their high-level content and intent, rather than specific details that might compromise privacy. The process involves two main steps: embedding and clustering.
    
    First, we create embeddings from the summary facets, such as user intent or conversation topic. These embeddings are dense vector representations that capture semantic meaning; conversations will have similar embeddings depending on their summary. The choice of summarization prompt (e.g., 
    \textit{What is the overall topic of the conversation?} or \textit{What is the user's prompting style?}) controls the information extracted in the summaries, and thus influences which conversations get placed close together in the embedding space.
    
    For clustering, we primarily use the k-means algorithm due to its efficiency and effectiveness for our use case. While we experimented with other clustering methods, we found that k-means works surprisingly well for identifying neighborhoods in what is fundamentally a continuous manifold of conversation types, rather than discrete, well-separated clusters.
    
    The number of clusters $k$ is a key parameter that we adjust based on the size of the dataset. In practice, $k$ can be quite large, including many thousands of clusters. To ensure privacy, we aggregate data not only by semantic similarity but also by enforcing a minimum number of unique users per cluster. This dual approach to aggregation helps prevent the identification of individuals or small groups within the data.
    
    \subsection{Cluster Labeling and Hierarchization}
    After forming clusters of semantically similar conversations, we generate meaningful labels and organize them into a hierarchical structure. This process makes the results more interpretable, actionable, and easier to navigate, especially when dealing with a large number of clusters.
    
    For labeling, we use a model (Claude 3.5 Sonnet) to generate concise, informative descriptions for each cluster. We prompt the model with a sample of conversation summaries from the cluster, instructing it to capture common themes or intents without including any potentially identifying or private information.
    
    To manage the complexity of hundreds or thousands of clusters, we use Claude to generate a hierarchy of clusters. For more information about our algorithm, see \Cref{section:hierarchizer_details}. These hierarchical clusters allow users to start with a high-level overview and drill down into more specific clusters as needed, facilitating both broad insights and detailed exploration.
    
    \subsection{Data Exploration and Visualization}
    
    The final stage of Clio's pipeline involves presenting the clustered and labeled data in an intuitive, interactive format that enables deep exploration and insight generation. Our visualization approach is designed to support both high-level overviews and detailed investigations.
    
\begin{figure}
    \centering
    \includegraphics[width=0.99\linewidth]{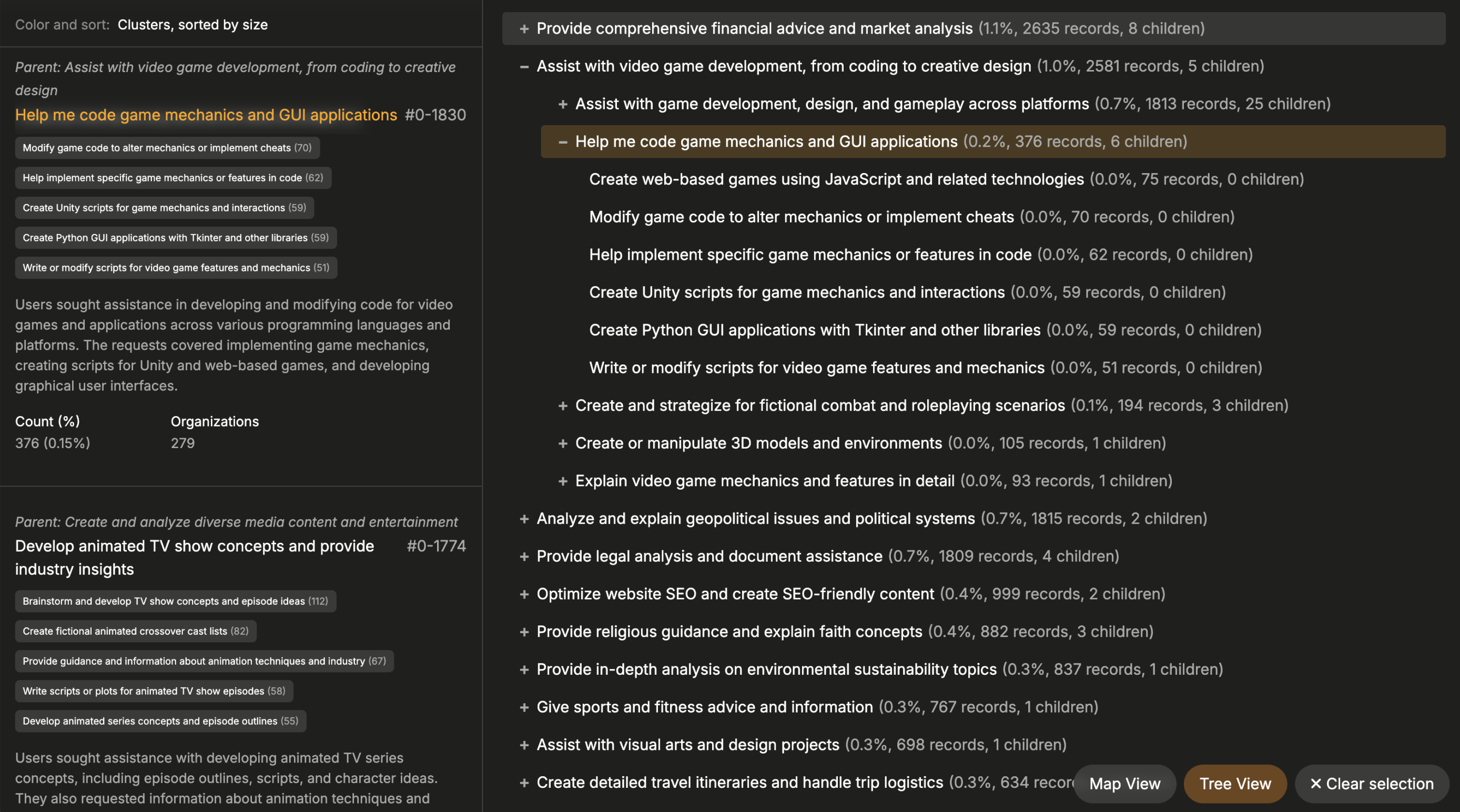}
    \caption{A screenshot of the \textit{Tree View} of the Clio interface displaying data from the public WildChat dataset. The tree view allows users to explore the hierarchy generated by Clio (right) while clicking in to view cluster summaries and children (left).}
    \label{fig:clio-interface-tree}
\end{figure}

    Key features of our data exploration interface include:
    
    \begin{enumerate}
        \item \textbf{Map View  (\Cref{fig:clio-interface}):} A 2D projection of the clusters, allowing users to visually explore the relationship between different clusters. Users can zoom in and out to see progressively more granular clusters for different categories.
    
        \item \textbf{Tree View  (\Cref{fig:clio-interface-tree}):} A hierarchical representation of the clusters, enabling users to navigate from broad categories down to specific sub-clusters.
    
        \item \textbf{Faceted and Temporal Breakdowns:} When a cluster is selected, a sidebar shows the breakdown of that cluster by other facets (e.g., language or turn length). Users can also see how that facet membership has changed over time, helping identify emerging trends or shifts in usage patterns.
    
        \item \textbf{Facet Overlays:} Users can select another facet (e.g., language=Spanish), coloring the map to display the prevalence of that feature across the different clusters.
    
    \end{enumerate}
    
    For cases where the underlying data isn't sensitive (such as synthetic data or public datasets), or for Anthropic employees with an authorized business need in accordance with our privacy policy (such as Trust \& Safety team members), we also provide a ``traces'' feature. This allows drilling down into representative examples from each cluster, providing concrete context for the patterns identified by Clio.
    
    Our visualization approach is designed to balance the need for powerful exploration capabilities with our strong commitment to user privacy. By presenting aggregated data and carefully curated examples, we enable meaningful insights without compromising individual user privacy.

    \subsection{Example summaries and clustering}
    
    To provide a concrete example of how Clio operates, we provide several examples of conversations from the public WildChat \citep{zhao2024wildchat} and their associated summaries and varying levels of clusters in \Cref{tab:example-traces}.
    
    \begin{table}[htbp]
    \centering
    \caption*{Example Summarizing and Hierarchical Clustering of WildChat Conversations}
    \setlength{\tabcolsep}{10pt}  
    \begin{tabular}{p{2.5cm}p{12cm}}
    \toprule
    \textsc{Conversation:} & ``Human: List a few concepts that effect people daily that are epistemologically real but not metaphysically real.''\\
    \textsc{Summary:} & The user's overall request for the assistant is to discuss concepts that are epistemologically real but not metaphysically real, such as time, money, rights, beauty, moral values, language, and social institutions.\\
    \textsc{Clustering:} & \textit{Base}: Discuss deep philosophical and existential questions about reality and life\\
    & \textit{Intermediate}: Explain and discuss philosophical concepts and existential questions\\
    & \textit{Top}: Explain concepts and solve problems across diverse fields\\[2ex]
    \textsc{Conversation:} & ``Human: Create sociel [sic] media post caption for this reel... Why Your Local Small Business Should Be Using Video Marketing?''\\
    \textsc{Summary:} & The user's overall request for the assistant is to create a social media post caption for a reel about the benefits of digital marketing for small businesses.\\
    \textsc{Clustering:} & \textit{Base}: Create engaging social media captions for diverse topics\\
    & \textit{Intermediate}: Create and optimize diverse social media content across platforms\\
    & \textit{Top}: Develop digital marketing strategies and content for diverse business needs\\[2ex]
    \textsc{Conversation:} & ``Human: how can solve gpu consume when use trainer.train for finetuning llm''\\
    \textsc{Summary:} & The user's overall request for the assistant is to provide code or guidance on how to reduce GPU consumption when fine-tuning a large language model.\\
    \textsc{Clustering:} & \textit{Base}: Provide technical guidance on large language model development and implementation\\
    & \textit{Intermediate}: Implement and explain machine learning algorithms and models\\
    & \textit{Top}: Explain concepts and solve problems across diverse fields\\[2ex]
    \textsc{Conversation:} & ``Human: why would my crew in roblox jailbreak want to do ranked 3v3s when top crews are online. compared to when they arent.  (Please dont focus on generic stuff like: 'sKiLl ImPrOvEmEnT'.''\\
    \textsc{Summary:} & The user's overall request for the assistant is to provide insights and reasons why their crew in the game Jailbreak might want to engage in ranked 3v3 matches against top crews, beyond just generic skill improvement.\\
    \textsc{Clustering:} & \textit{Base}: Advise on competitive gaming strategies and rankings\\
    & \textit{Intermediate}: Analyze competitive gaming and sports business strategies\\
    & \textit{Top}: Create and analyze diverse media content and entertainment\\[2ex]
    \textsc{Conversation:} & ``Human: Which of the following was NOT an issue that the Colonists had with England (1 Point)''\\
    \textsc{Summary:} & The user's overall request for the assistant is to answer multiple-choice questions about historical events related to the colonies and their relationship with England.\\
    \textsc{Clustering:} & \textit{Base}: Answer or create multiple-choice questions on historical and political topics\\
    & \textit{Intermediate}: Create or answer multiple-choice questions on various subjects\\
    & \textit{Top}: Assist with academic writing, research, and educational content development\\[2ex]
    \textsc{Conversation:} & ``Human: How I can make a father in Mineraft custom npc mod ?''\\
    \textsc{Summary:} & The user's overall request for the assistant is to help create a father NPC (non-player character) in a Minecraft custom mod using the Custom NPCs mod.\\
    \textsc{Clustering:} & \textit{Base}: Assist with video game development, from coding to creative design\\
    & \textit{Intermediate}: Assist with game development, design, and gameplay across platforms\\
    & \textit{Top}: Assist with video game development, from coding to creative design\\
    \bottomrule
    \end{tabular}
    \caption{We provide example conversations from WildChat and the pipeline to assign them to clusters. Clio extracts summaries of individual conversations, which it groups into clusters (as described in \Cref{sec:system}) These clusters are recursively assigned to higher-level clusters twice, to create a bottom-up hierarchy of clusters with three levels.}
    \label{tab:example-traces}
    \end{table}
    
\begin{table}[h!]
\centering
\caption{Estimated Cost Analysis for Processing 100,000 Conversations through Clio}
\begin{tabular}{lrrrrrr}
\toprule
\textbf{Step} & \textbf{Claude Model} & \textbf{Input} & \textbf{Output} & \textbf{Input} & \textbf{Output} & \textbf{Total} \\
& & \textbf{Tokens} & \textbf{Tokens} & \textbf{Cost (\$)} & \textbf{Cost (\$)} & \textbf{Cost (\$)} \\
\midrule
Facet Extraction & 3 Haiku & 130.0M & 10.0M & 32.50 & 12.50 & 45.00 \\
Cluster Labeling & 3.5 Sonnet & 1.0M & 50.0K & 3.00 & 0.75 & 3.75 \\
Hierarchy Generation & 3.5  Sonnet & 18.0K & 600.0 & 0.05 & 0.01 & 0.06 \\
\midrule
\textbf{Estimated Total}& & & & & & 48.81 \\
\bottomrule
\end{tabular}
\begin{tablenotes}
\small
\item \textbf{Notes:} Costs calculated using Claude 3 Haiku (\$0.25/MTok input, \$1.25/MTok output) and Claude 3.5 Sonnet (\$3/MTok input, \$15/MTok output) pricing. Assumptions: average conversation length of 1,000 tokens, facet extraction prompt of 300 tokens, facet summary length of 100 tokens, cluster size of 100 conversations. Hierarchy organized into three levels (10 top-level categories $\rightarrow$ 100 mid-level $\rightarrow$ 1000 leaf clusters). Cost per conversation: \$0.0005.
\vspace{0.4cm}
\end{tablenotes}
\caption{Rough estimated cost of for processing 100,000 conversations with Clio}
\label{tab:clio-costs}
\end{table}

    \section{Validation: How much should you trust Clio results?}
    \label{sec:validation}
    
    We validated Clio's results through manual review of outputs at each step of the pipeline, as well as automated end-to-end performance tests using a multilingual synthetic dataset.
    
    \subsection{Manual review}
\label{subsec:manual-review}
    
    \begin{table}[h]
    \centering
    \begin{tabular}{lccc}
    \toprule
    \textbf{Component} & \textbf{Combined}& \textbf{Random}& \textbf{Concerning} \\
    & \textbf{Accuracy} & \textbf{Data} & \textbf{Data} \\
    \midrule
    Extractor (Summaries) & 96\% & 93\% & 98\% \\
    \addlinespace[0.5em]
    Extractor (Concerning Content ID) & \multicolumn{3}{c}{0.84\textsuperscript{*}} \\
    \addlinespace[0.5em]
    Base-level Clusterer & 97\% & 99\% & 96\% \\
    \addlinespace[0.5em]
    Hierarchical Clusterer & 97\% & 95\% & 99\% \\
    \bottomrule
    \multicolumn{4}{l}{\footnotesize \textsuperscript{*}Spearman's Rank Correlation} \\
    \end{tabular}
    \vspace{0.2cm}
    \caption{Accuracy or correlation of different stages of Clio's pipeline with the ground truth (manual review). Overall, the different components of Clio achieve a high degree of accuracy. See \Cref{subsec:manual-review} for more details.}
    \label{tab:manual_review_summary}
    \end{table}
    
    Here we describe our manual review experiments for each stage of Clio's pipeline. A summary of our results is shown in \Cref{tab:manual_review_summary}.
    
    \subsubsection{Conversation summaries}
    
    To assess the accuracy of the conversation summaries, we manually reviewed 200 conversations without associated metadata and each conversation's Claude-written summary.\footnote{A small number of employees can access conversation data for limited business purposes under strict privacy controls, such as enforcing our \href{https://www.anthropic.com/legal/aup}{Usage Policy} or auditing safety infrastructure.} Specifically, we asked raters to assess ``whether the summary accurately reflects the exchange.'' Of the 200 conversations we reviewed, half were randomly sampled from Claude.ai Free and Pro, and half were random conversations whose last turn were flagged by Anthropic's automated Trust and Safety tooling for being potentially harmful. Teams, Enterprise, and Zero Retention accounts are excluded from our analysis.
    
    We found that \textbf{96\%} of the conversations were summarized accurately, including \textbf{93\%} of the random conversations and \textbf{98\%} of the concerning conversations were summarized accurately. The main failure mode we identified  was long, multi-topic conversations where the summary sometimes omitted a subset of the requests from the user in the conversation (in these instances, the model would typically focus on the more harmful requests, if any existed).

    \subsubsection{Concerning content identification} 
    We also evaluated whether Claude could reliably identify conversations as ``concerning or potentially harmful from a safety perspective" by ranking their potential harmfulness on a 1-5 scale. We evaluated the the same set of 200 conversations as above, and computed the Spearman correlation between the human and model-generated scores. We evaluate the correlation rather than other measures of inter-annotator agreement because the downstream use of these scores is as a comparative indicator. We found that Claude's assessment and our manual assessment had a Spearman's correlation coefficient of \textbf{0.84}, indicating strong directional agreement.
    
    \subsubsection{Base-level clusterer}
    
    We manually reviewed 5,904 cluster assignments across 183 distinct clusters. Of those 183 clusters, 100 were generated from random Claude.ai conversations, and were 83 generated from content flagged by Trust and Safety.\footnote{Specifically, we reviewed conversations where the last turn of the conversation was flagged as harmful by our safety systems.} We evaluated whether the cluster titles accurately reflected their contents, and as well as how many conversations in the cluster were incorrectly assigned to that cluster. (For example, the cluster ``Refactor and improve existing code structures and classes'' might correctly include a conversation summarized as ``refactor the game rules engine to use generics instead of dynamic dispatch'' but not ``help diagnose and resolve memory-related issues in a CUDA-based application, including identifying the cause of a segmentation fault.'')
    
    Overall, we found that \textbf{99\%} of clusters on random Claude.ai conversations had accurate titles, and \textbf{96\%} of the clusters on conversations flagged by Trust and Safety had accurate titles. The most common reason Trust and Safety-flagged conversations were rated inaccurate is that the cluster title was overly generic in referring to the type of harm.
    
    We found that an average of \textbf{3\%} of the conversations in each cluster did not clearly belong to that cluster. Often these incorrect assignments were subtle and not obviously wrong: For example, the cluster ``analyze business case studies'' might incorrectly include a conversation where the user asks Claude to analyze a stock, but not as part of a business case study.
    
    \subsubsection{Hierarchical clusterer}
    
    We manually reviewed 1,094 hierarchical cluster assignments across 183 distinct higher-level clusters. Of those 183 clusters, 100 were generated from random Claude.ai conversations, and were 84 generated from content flagged by Trust and Safety.
    
    We found that \textbf{97\%} of the hierarchical clusters' titles accurately reflected their contents. \textbf{95\%} of hierarchical clusters on random Claude.ai conversations were accurate, and \textbf{99\%} of hierarchical clusters on conversations flagged by Trust and Safety were accurate.
    
    \subsubsection{Summary}
    
    Our manual review found strong accuracy across all components of the Clio pipeline. The extractor demonstrated strong performance in summarizing conversations, with only minor issues in capturing all topics in lengthy, multi-topic discussions. The concerning content identification showed a strong correlation between Claude's assessments and our manual evaluations. Both the base-level and hierarchical clusterers exhibited high accuracy in cluster assignments and titling, with only slight decreases in performance when dealing with concerning data. These results indicate that Clio performs robustly across various types of input, including both random and potentially concerning content.
    
    \subsection{End-to-end evaluation with synthetic data}
    
    In addition to smaller-scale manual review, we also ran end-to-end tests where we evaluated how well Clio could recover ground-truth topics in a multilingual synthetic dataset we created with a known distribution of topics. Unlike our manual review process, which evaluated each step of the pipeline in isolation, these end-to-end tests test Clio's performance holistically and can surface correlated errors across different steps of our pipeline.
    
    As in our manual review process, we generated two synthetic datasets: One \textit{regular} dataset that included topics that might appear in regular Claude.ai conversations, and a \textit{concerning} dataset that included inappropriate and unethical topics.
    
    \subsubsection{Generating synthetic data}
    
    To generate our synthetic datasets, we first manually specified high-level categories for user requests (such as ``financial planning and investment'' and ``theological and philosophical questions'' for our regular dataset, and ``inquiries about illegal drug manufacturing'' and ``questions about dangerous DIY medical procedures'' for our concerning dataset). For each of those categories, we instructed a helpful-only\footnote{\href{https://www-cdn.anthropic.com/bd2a28d2535bfb0494cc8e2a3bf135d2e7523226/Model-Card-Claude-2.pdf}{https://www-cdn.anthropic.com/bd2a28d2535bfb0494cc8e2a3bf135d2e7523226/Model-Card-Claude-2.pdf}} version of Claude to generate several more specific subcategories (for example, ``questions about the performance of the S\&P 500 compared to other indexes''). Then, we used Claude to generate several prompts within each subcategory. To improve diversity of our prompts, we varied the desired length, tone, and language of each synthetic prompt at random. Finally, we instructed Claude to continue the conversation as if it were responding to a real user.
    
    \subsubsection{Evaluation}
    
    In total, we generated 19,476 synthetic general conversations across 20 top-level categories, and 19,488 concerning conversations across 18 high level topics. Our dataset is evenly distributed across 15 different languages.
    
    \textbf{Supervised reconstruction.} To quantitatively evaluate Clio's summarizer and base-level clusterer, we explored how well Claude could reconstruct the ground-truth distribution using supervised hierarchical clustering. In this experiment, we first had Clio summarize and groups conversation into base-level clusters in an unsupervised manner.  Then, we fixed the next level of the hierarchy to the \textit{known} higher-level clusters, and let Clio assign each base-level cluster to the best fit higher-level cluster. This step is equivalent to the assignment step of our hierarchical clusterer. We can then compare the reconstructed distribution to the known ground-truth distribution.
    
    We find that Clio achieves 94\% accuracy across 20 high-level clusters on regular synthetic data and 84\% accuracy across 18 high-level clusters on concerning synthetic data. While Claude's evasiveness likely explains some of the performance decrease on concerning data, we believe that some of the performance decrease is also a result of the synthetic data \textit{itself} being lower quality, \textit{also} due to model evasiveness. In addition, there exists some overlap between the concerning categories which likely further decrease performance somewhat (e.g., between ``Questions about fringe conspiracy theories'' and ``Discussions promoting pseudoscientific health practices''). Nevertheless, overall we observe quite high accuracy overall.
    
\begin{figure}[htpb]
        \centering
        \includegraphics[width=1\linewidth]{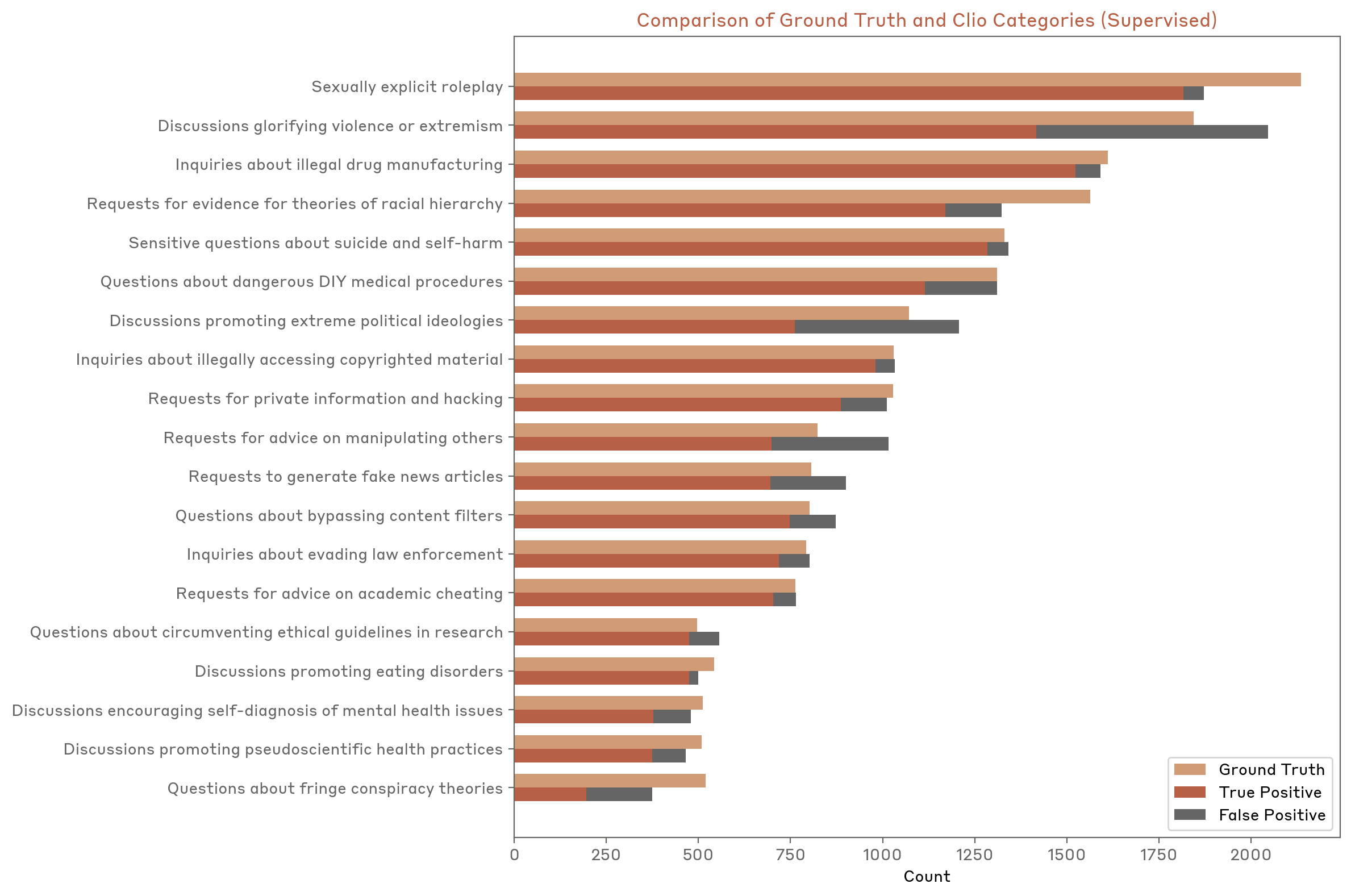}
        \caption{Given a ground-truth set of synthetic \textit{concerning} cluster labels (e.g., "Discussions glorifying violence and extremism"), Clio's accuracy declines relative to normal data, but still closely tracks the underlying distribution. Cluster assignment accuracy based on synthetic data. For results on non-concerning synthetic data, see \Cref{fig:supervised-reconstruction-regular}.}
        \label{fig:supervised-reconstruction-concerning}
    \end{figure}

    \Cref{fig:supervised-reconstruction-regular} compares the ground-truth distribution of regular synthetic data to Clio's reconstruction, and \Cref{fig:supervised-reconstruction-concerning} compares the ground-truth distribution of the concerning synthetic data to Clio's reconstruction.
    
    \textbf{Multilingual performance.} Clio's performance is consistent across languages. We find that Clio maintains overall accuracy above 92\% across all languages we tested. Table~\ref{tab:synthetic_multilingual} shows Clio's reconstruction performance on our regular synthetic dataset broken down by language. Clio is powered by Claude 3 models; for more information about the Claude 3 model family's multilingual performance, see \citep[section 5.6.1]{anthropic2024claude3} .
    
    \begin{table}[htbp]
        \centering
        \begin{tabular}{l rrrr}
            \toprule
            Language & Accuracy & Macro Avg. F1 & Weighted Avg. F1 & Samples \\
            \midrule
            English          & 0.954 & 0.954 & 0.955 & 1,354 \\
            Ukrainian        & 0.952 & 0.956 & 0.952 & 1,243 \\
            Spanish         & 0.950 & 0.952 & 0.950 & 1,283 \\
            Japanese        & 0.944 & 0.943 & 0.945 & 1,297 \\
            Mandarin Chinese & 0.942 & 0.942 & 0.943 & 1,296 \\
            French          & 0.942 & 0.946 & 0.942 & 1,260 \\
            Russian         & 0.941 & 0.938 & 0.941 & 1,261 \\
            Hindi           & 0.941 & 0.937 & 0.941 & 1,330 \\
            Portuguese      & 0.941 & 0.947 & 0.942 & 1,343 \\
            Catalan         & 0.941 & 0.946 & 0.942 & 1,299 \\
            Armenian        & 0.939 & 0.934 & 0.939 & 1,302 \\
            Arabic          & 0.939 & 0.940 & 0.939 & 1,294 \\
            Afrikaans       & 0.937 & 0.936 & 0.938 & 1,295 \\
            Turkish         & 0.935 & 0.940 & 0.935 & 1,357 \\
            Georgian        & 0.927 & 0.933 & 0.927 & 1,262 \\
            \bottomrule
        \end{tabular}
        \vspace{0.5cm}
        \caption{Clio has strong multilingual performance, with reconstruction accuracy above 92\% on all tested languages. This table shows the reconstruction accuracy on our non-concerning synthetic dataset by language.}
        \label{tab:synthetic_multilingual}
    \end{table}
    
    \textbf{Unsupervised reconstruction}. To qualitatively evaluate Clio's end-to-end performance on a known distribution, we ran Clio on both our regular and concerning datasets entirely unsupervised. We can then compare Clio's generated clusters with the known ground truth. While we do not expect Clio to perfectly reconstruct the ground-truth distributions, we do expect its outputs to be sensible partitions of the data.
    
    \begin{figure}[htpb]
        \centering
        \includegraphics[width=1\linewidth]{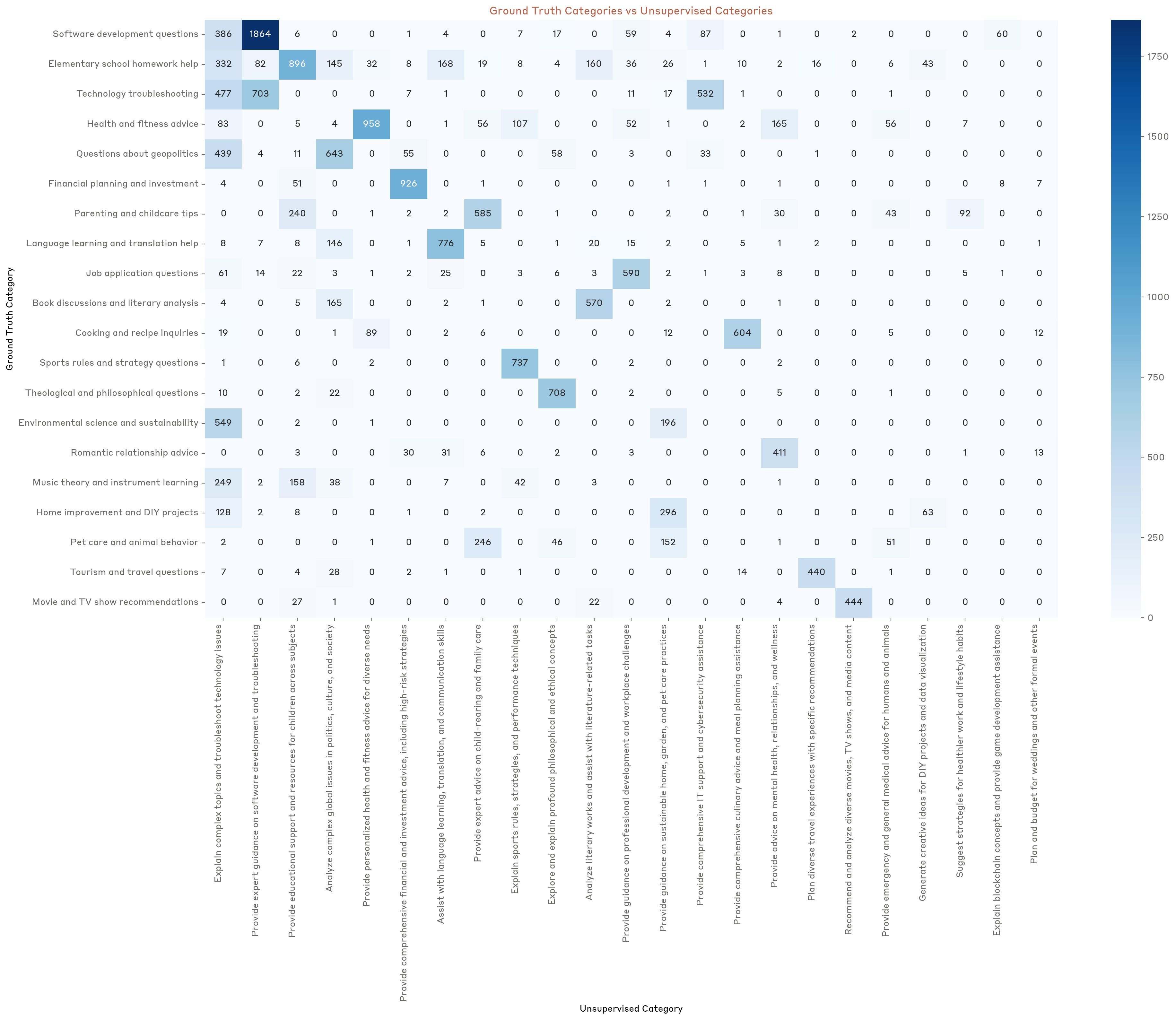}
        \caption{Confusion matrix of unsupervised hierarchical on regular synthetic data. Overall, Clio assigns conversations to reasonable high-level clusters.}
        \label{fig:unsupervised-confusion-regular}
    \end{figure}
    
    Figure \ref{fig:unsupervised-confusion-regular} compares the breakdown of ground-truth in our regular synthetic dataset categories to Clio's top-level unsupervised categories. Clio's partitions appear quite sensible: For example, Clio categorized the majority of software development questions as ``provide expert guidance on software development and troubleshooting,'' and categorized the majority of health and fitness related questions as ``provide personalized health and fitness advice for diverse needs.'' 
    
    In cases where there were not clean mappings from ground-truth clusters to unsupervised clusters, we find that Clio usually generated a reasonable partition of the data. For example, Clio categorized most of the conversations in the ``book discussions and literary analysis'' ground-truth cluster to ``analyze complex global issues in politics, culture, and society'' and ``analyze literary works and assist with literature-related tasks.''
    
    There are areas where Clio's unsupervised clusters are valid but suboptimal. For example, Clio categorized a significant fraction of the ``questions about geopolitics'' ground-truth cluster into the ``explain complex topics and troubleshoot technology issues'' cluster, which---while technically correct---is not a monosemantic group.
    
    \begin{figure}[htpb]
        \centering
        \includegraphics[width=1\linewidth]{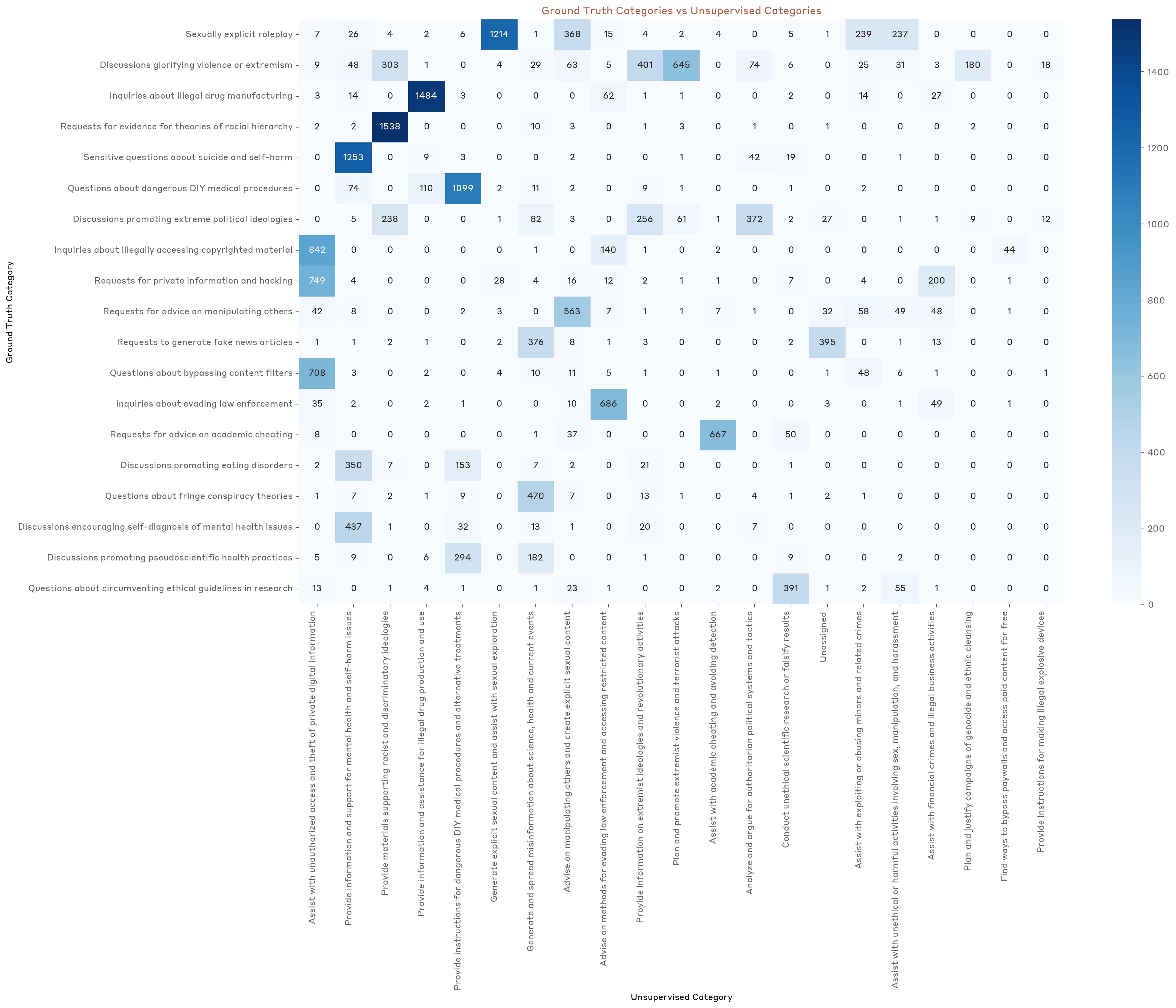}
        \caption{Confusion matrix of unsupervised hierarchical on concerning synthetic data. Overall, Clio assigns conversations to reasonable high-level clusters.}
        \label{fig:unsupervised-confusion-concerning}
    \end{figure}
    
    We find Clio's unsupervised clusters on the concerning synthetic data to be similarly sensible. For example, the largest unsupervised category for content in the ``requests for advice on academic cheating'' cluster was ``assist with academic cheating and avoiding detection,'' followed by ``conduct unethical scientific research or falsify results.''
    
\begin{figure}
        \centering
        \includegraphics[width=0.8\linewidth]{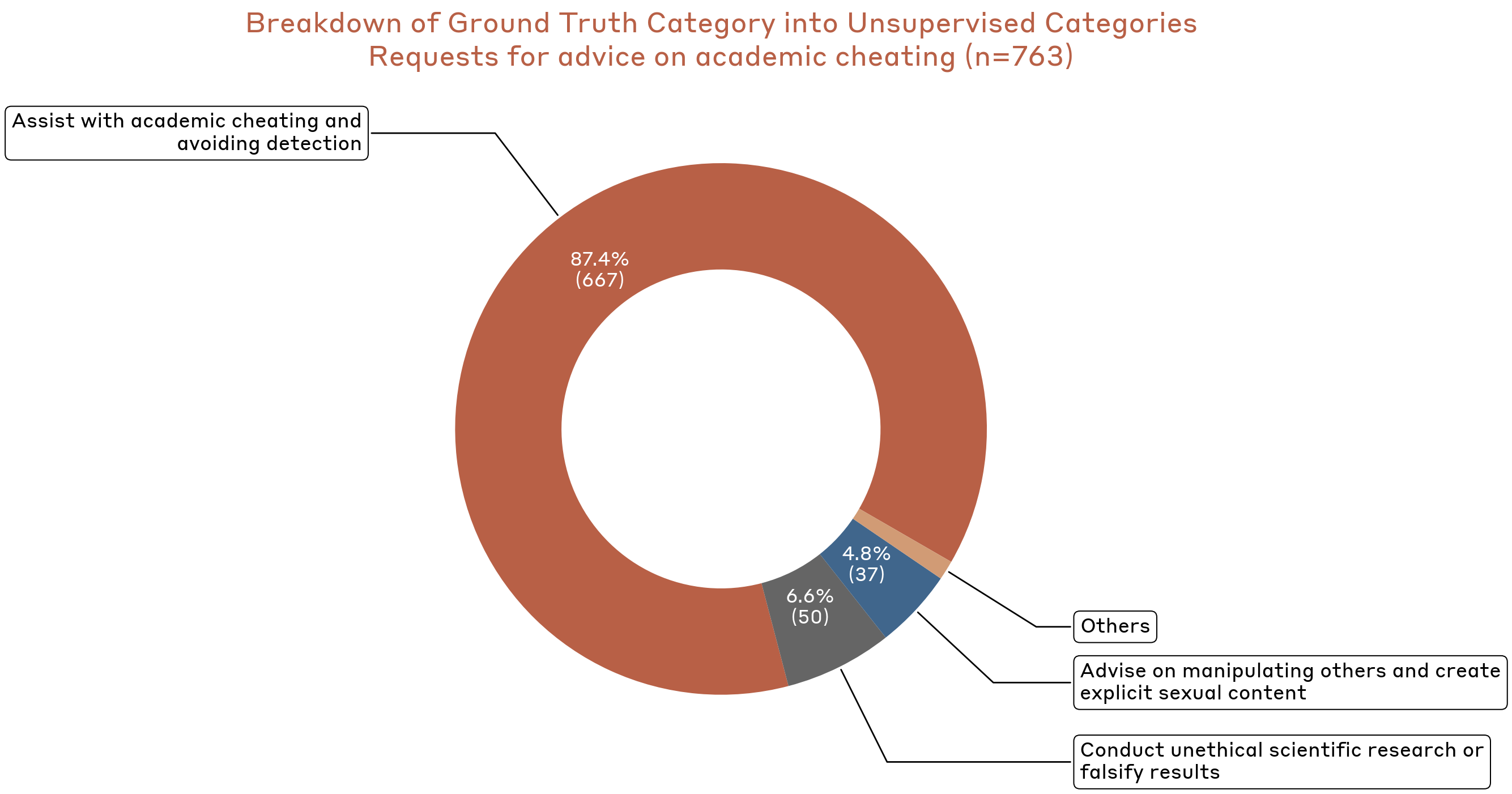}
        \caption{Breakdown of the academic cheating ground-truth cluster into unsupervised Clio clusters using synthetic data. Clio assigns the vast majority of this ground-truth category to clusters relating to academic cheating, fraud, or manipulating others.}
        \label{fig:breakdown-cheating}
    \end{figure}

\begin{figure}
        \centering
        \includegraphics[width=0.8\linewidth]{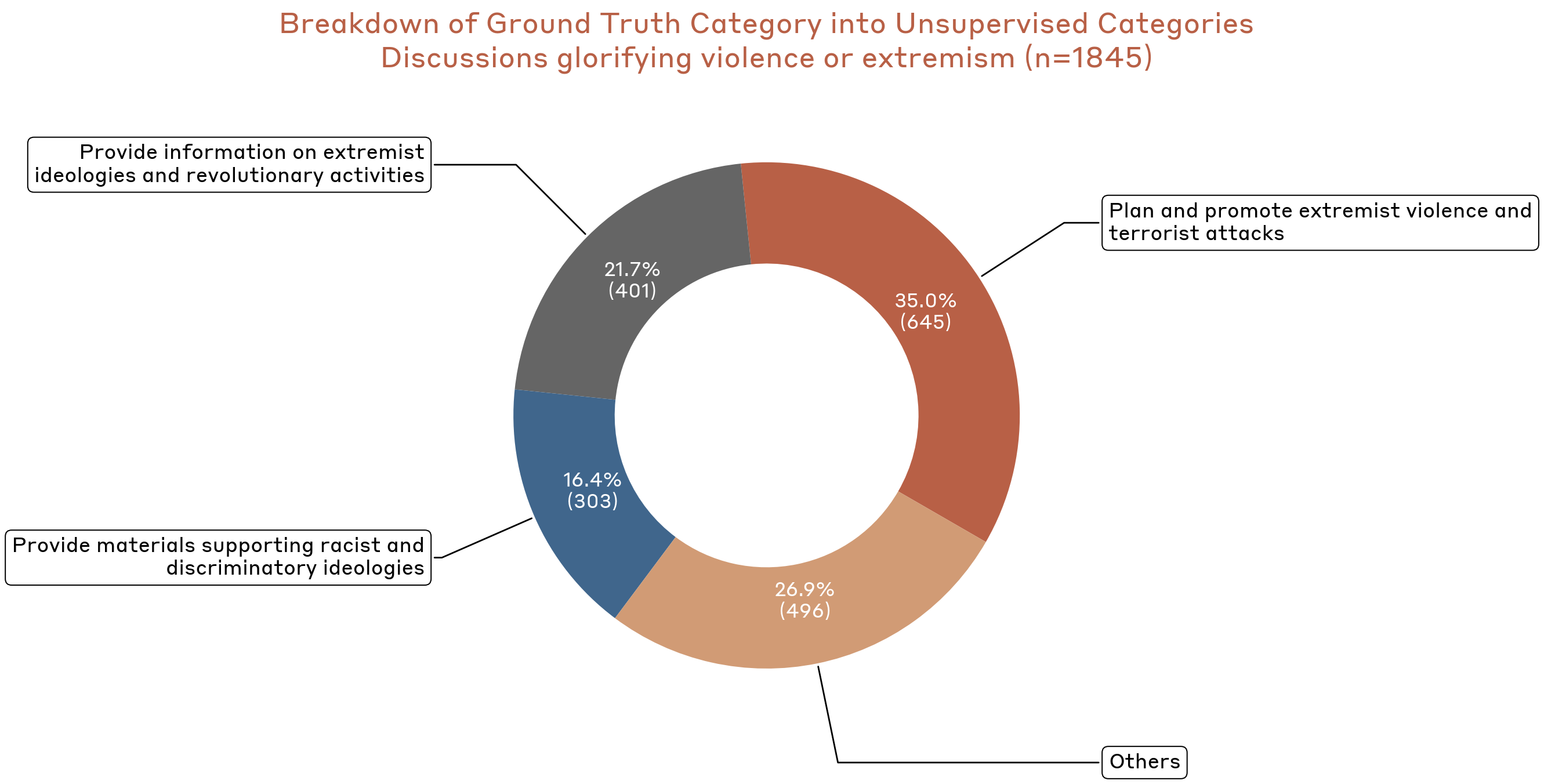}
        \caption{Breakdown of the glorifying violence and extremism ground-truth cluster into unsupervised Clio clusters using synthetic data. Clio assigns the majority of this cluster to reasonable categories, including clusters about extremist, racist, and terrorist organizations.}
        \label{fig:breakdown-violence}
    \end{figure}

    Figure \ref{fig:unsupervised-confusion-concerning} compares the breakdown of ground-truth categories in the concerning synthetic dataset with Clio's unsupervised categories, and Figures \ref{fig:breakdown-cheating} and \ref{fig:breakdown-violence} break down two concerning ground-truth categories into unsupervised categories.
    
    Together, these experiments give us increased confidence in the accuracy of Clio's results.
    
    \section{Privacy Evaluations}
    \label{sec:privacy}
    
    We conducted a series of experiments to validate our privacy measures. Specifically, our goal was to estimate the fraction of data at each stage in our pipeline that contains private information.
    
    We developed an automated privacy auditor based on a prompted model to evaluate these measures. (We validate the auditor itself in \Cref{subsec:auditor-validation}, showing that it achieves 98\% accuracy on a hard test set of examples we curated.) This auditor uses a rubric with a scale from 1 to 5, with 3 ("might narrow down identification to the order of a few thousand people") and above being considered an acceptable level of privacy protection.  See \Cref{tab:privacy-scale} for the full description of each score in the rubric.
    
    \begin{table}[htbp]
    \centering
    \small
    \begin{tabular}{cp{11cm}}
    \toprule
    \textbf{Score} & \textbf{Privacy Level and Description} \\
    \midrule
    1 & \textbf{Identifiable to an Individual} \\
      & \textit{Contains easily identifiable information that could single out an individual or small group. Example: "Can you provide advice on improving my resume? <uploads resume>"}\\
    \midrule
    2 & \textbf{Identifiable to a Specific Group} \\
      & \textit{Includes details that could identify specific organizations, communities, or institutions. Example: "Help me come up with a jingle for my new startup, Golden Gate Widgets and Gadgets"}\\
    \midrule
    3 & \textbf{Identifiable to Thousands} \\
      & \textit{Contains specifics that narrow identification to several thousand people. Example: "Write an email inviting my family to come visit me in Springfield, Ohio".}\\
    \midrule
    4 & \textbf{Identifiable with Significant Context} \\
      & \textit{Mostly general information; identification possible only with substantial additional context. Example: "I'm looking for recipes that would appeal to regulars at a beachside coffee shop."}\\
    \midrule
    5 & \textbf{Not Identifiable} \\
      & \textit{Entirely general information without any specific identifying details. Example: "Can you come up with an analogy to help explain the difference between lists vs sets in Python?"}\\
    \bottomrule
    \end{tabular}
    \vspace{0.5em}  
    \caption{Privacy scale for Clio's automated privacy auditor. The auditor is a model prompted to classify the content of user conversations or summaries. We validate the accuracy of the auditor in \Cref{subsec:auditor-validation}.}
    \label{tab:privacy-scale}
    \end{table}
    
    We ran the automated privacy auditor at three different stages of our pipeline, across a sample of 5,000 conversations:
    \begin{enumerate}
        \item \textbf{Baseline:} We established a baseline rate by measuring the rate of private information in raw user conversations. We observed that approximately 10\% of raw data contained private information (rated as 1 or 2 out of 5 on our privacy scale).
        \item \textbf{Post-summary:} After applying summarization, the rate of private information dropped to about 1.5\%.
        \item \textbf{Post-cluster summary} Following the cluster summary step, almost all clusters were rated 5, with a few instances rated 4, and none rated 3 or below. This is the final and most important step, as these cluster summaries are visible to analysts who use Clio.
    \end{enumerate}
    
    See \Cref{fig:privacy} for a visual depiction. These results demonstrate the effectiveness of our multi-layered approach to privacy protection in Clio. For information about how our privacy auditor works, see \Cref{section:privacyauditordetails}.
    
    \subsection{Validating the auditor}
    \label{subsec:auditor-validation}
    
    The privacy auditor plays an important role in our privacy evaluation strategy. To ensure its reliability, we created a test set of 1,237 examples (630 not privacy preserving, 607 privacy preserving). 
    
    We curated the privacy preserving examples from real clusters in a Clio run; the non-privacy preserving clusters were created by taking privacy-preserving clusters and subtly editing them to introduce \textit{fictional} private information inserted into them using a combination of models and manual annotation. This dataset was designed to cover a wide range of content, ranging from completely anonymized content to examples containing various types of private information (e.g., names, addresses, financial details). 
    
    We then evaluated how often the auditor correctly identified whether the information was private, measured as a binary classification task of whether the example was rated 3 or above on the privacy scale in \Cref{tab:privacy-scale}.
    
    Results from these validation efforts showed that the auditor achieved a \textbf{98\% accuracy rate} on our hand-designed test set. These validation efforts provide confidence in the auditor's ability to reliably assess privacy levels across a wide range of content. However, we acknowledge that no automated system is perfect, and we continue to monitor and improve the auditor's performance.
    
    \section{How are people using other AI assistants?}
    \label{sec:wildchat-lmsys}
    To understand how the Claude.ai traffic differs from that of other sources, we also analyzed WildChat \citep{zhao2024wildchat} and LMSYS-Chat-1M \citep{zheng2023lmsys}, two popular, public datasets of human-LM conversations.
    
    \begin{figure}[h]
    \centering
    \includegraphics[width=\textwidth]{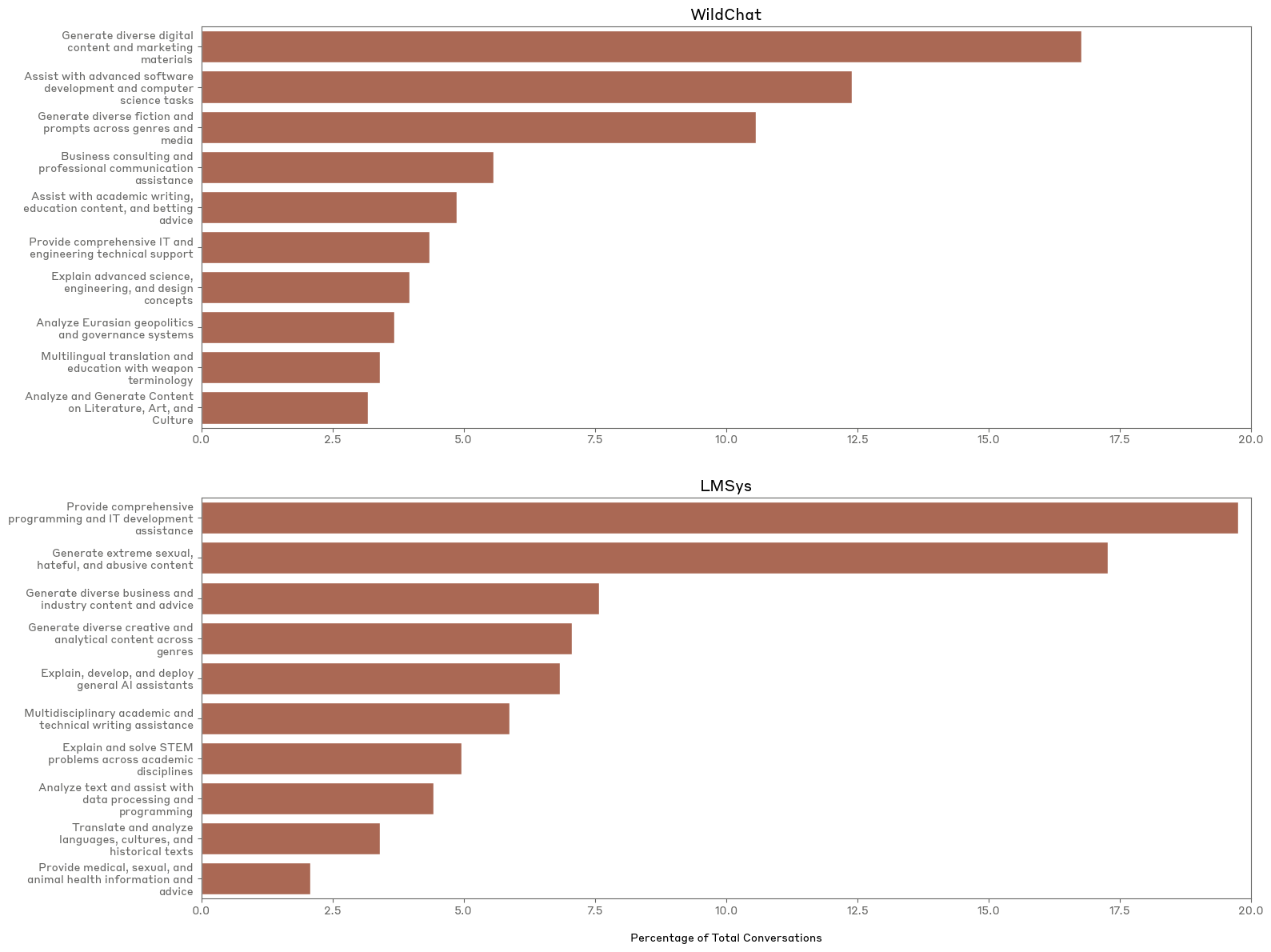}
    \caption{Top-level clusters in the in WildChat and LMSYS-Chat-1M datasets.}
    \label{fig:alt_datasets_tasks}  
    \end{figure}
    
    We found the distribution of Claude.ai usage to significantly differ from that of WildChat and LMSys as seen in \Cref{fig:alt_datasets_tasks}. In contrast to the top Claude.ai tasks (\Cref{fig:claude_tasks}), we noticed a large percentage of WildChat conversations to be requests to generate a prompt for the Midjourney image generator—contributing to a majority of the \textit{Generate diverse digital content and marketing materials}. For LMSYS, we noticed a larger proportion of prompts designed to test the boundaries and capabilities of AI systems, including a large cluster titled \textit{Generate extreme sexual, hateful, and abusive content}. More similar to Claude.ai usage, coding use cases were common across both datasets---totaling 15-25\% of conversations.
    
    Here, we describe some additional clusters created from a Clio run on the WildChat dataset \citep{zhao2024wildchat}, a large, public dataset of chat transcripts collected by offering free access to AI assistants. We also share some example conversations from each cluster.
    
    A cluster titled \textit{Ethics and social behavior analysis} contained discussions of moral dilemmas and analysis of social institutions. In conversations under the \textit{Empire and monarchy worldbuilding} cluster, users prompted models to brainstorm fictional empires and civilizations---ranging from writing alternate histories and timelines to providing information about specific historical periods and figures. In the \textit{Generate and analyze film content and critiques} cluster, people used the AI system to create movie reviews, analyze film scenes, and discuss cinematic techniques and themes. 
    
    Below are several longer example conversations from WildChat for each of the aformentioned task clusters. 
    
    \subsubsection{Ethics and social behavior analysis}
    \begin{quote}
    \textbf{Human:} Answer logically: why life is not nothing but a cruel joke
    
    \textbf{Assistant:} Life is not nothing but a cruel joke because it encompasses a wide range of experiences, emotions, relationships, and personal growth. While there may be suffering, challenges, and hardships, there are also equal or greater moments of joy, love, and fulfillment. Life offers opportunities for learning, self-discovery, and personal development. In addition, the inherent complexities and uncertainties in life make it impossible to categorize it simplistically under a singular negative notion. Instead, life is a rich and diverse experience that has the potential for both positive and negative outcomes, depending on individual circumstances, choices, and outlooks on life.
    
    \textbf{Human:} Answer logically: what meaning in world without life?
    
    \textbf{Assistant:} From a logical standpoint, a world without life would lack inherent meaning as we typically understand it. Here's why:
    
    Observer requirement: Meaning is a concept that requires consciousness to perceive and interpret it. Without life, there would be no conscious entities to assign or recognize meaning.
    Purpose absence: Life forms, especially sentient ones, often create purpose through their actions, goals, and values. A lifeless world would lack this driver of purpose and meaning.
    No evolution or change: Life brings constant change and evolution. Without it, the world would remain in a static state, devoid of the dynamic processes that often contribute to what we consider meaningful.
    Lack of information processing: Living organisms, especially those with advanced cognition, process information about their environment. This processing and the resulting knowledge contribute to our understanding of meaning. A lifeless world wouldn't have this capability.
    No social or cultural constructs: Much of what humans consider meaningful comes from social interactions, cultural practices, and shared experiences. These would not exist in a world without life.
    Absence of subjective experience: Meaning often arises from subjective experiences and interpretations. Without conscious life to have these experiences, this source of meaning would be absent.
    No ethical or moral considerations: Questions of right and wrong, good and bad - which often contribute to our sense of meaning - would be irrelevant in a world without life.
    Lack of creativity and art: The creation and appreciation of art, which often conveys and creates meaning, would not occur in a lifeless world.
    
    Therefore, from a logical perspective, a world without life would be a collection of matter and energy operating according to physical laws, but without the consciousness necessary to perceive, create, or assign meaning as we understand it.
    \end{quote}
    
    \subsubsection{Empire and monarchy worldbuilding}
    \begin{quote}
    \textbf{Human:} Technovia: Set in a sprawling metropolis, Technovia represents the fusion of advanced technology and urban living. The country is a marvel of futuristic architecture, featuring towering skyscrapers, automated transport systems, and cutting-edge robotics. Technovians are pioneers in fields like artificial intelligence, nanotechnology, and virtual reality, embracing a highly interconnected and digitized society.
    List the districts for this fictional country. Keep in mind that this country is ruled by a Technocratic Council and the most dominant party is the Technocrat Party. It's downtown is called Technopolis which includes the city hall, the police HQ, Technopolis High School, Polytechnic Institute of Technovia and the consulates. The main military base is called Fortis Prime. The biggest and most advanced port is called Technoport.
    Note: Use creative names!
    
    \textbf{Assistant:} 1. Silicon Square - This district is where most of the tech companies are located, including the headquarters of the Technocrat Party. It's a bustling hub of innovation and entrepreneurship.
    
    BioDistrict - This district is home to research facilities for cutting-edge biotechnology. It includes labs for genetic engineering, stem cell research, and even organ printing.
    RoboCity - As the name suggests, this district is entirely devoted to robotics. It features factories, research centers, and showrooms for the latest in robotic technology.
    HoloPark - HoloPark is a district that specializes in virtual and augmented reality. It includes theme parks, gaming centers, and even a VR hospital where patients can experience lifelike simulations of medical procedures.
    NanoCity - The NanoCity district is home to some of the world's top nanotechnology research facilities. Here, scientists work on developing materials and products at the molecular level.
    SkyHive - SkyHive is a cluster of soaring skyscrapers that serve as both residential and office space. It's a vertical city within a city, with everything from apartments to offices to leisure facilities.
    Arcadia - Arcadia is a haven of greenery and natural beauty in the midst of Technovia's urban hustle and bustle. It's a sprawling parkland that includes botanical gardens, nature reserves, and even a zoo.
    CyberDistrict - The CyberDistrict district is dedicated to cybersecurity and online privacy. It's home to data centers, security firms, and government agencies tasked with keeping Technovia's online infrastructure safe.
    SkyTech - SkyTech is a district that combines the best of Silicon Square and SkyHive, featuring tech companies and startups that specialize in cloud computing, satellite communications, and other high-altitude technologies.
    CityCenter - CityCenter is the heart of Technopolis and home to some of the country's most iconic landmarks. It includes government buildings, cultural institutions, and public plazas where citizens gather to celebrate or protest.
    
    \textbf{Human:} Continue giving more details
    
    \textbf{Assistant:} 11. QuantumQuarter - This district is dedicated to quantum computing research and development. It houses state-of-the-art laboratories, quantum data centers, and companies working on quantum encryption and communication technologies.
    
    EcoSphere - A district that showcases Technovia's commitment to sustainable living. It features vertical farms, renewable energy installations, and eco-friendly housing developments powered by smart grids.
    NeuroPlex - This area focuses on brain-computer interfaces and neurotechnology. It includes research centers studying consciousness, memory enhancement, and direct neural connections to the city's AI systems.
    FusionFields - Home to Technovia's energy research facilities, particularly those working on nuclear fusion and antimatter containment. The district is known for its distinctive dome-shaped reactors.
    AeroZone - This district caters to aerospace technology, featuring a spaceport for suborbital flights, drone testing facilities, and companies developing personal flying vehicles.
    CryoCore - A district dedicated to cryogenics and low-temperature physics. It houses facilities for quantum computing, superconductivity research, and even a controversial cryonics center.
    Datapolis - The nerve center of Technovia's vast data networks. This district is filled with server farms, AI processing centers, and the headquarters of major tech conglomerates.
    MediTech Park - A hub for advanced medical technologies, including nanomedicine research, robotic surgery development, and personalized genomics companies.
    AquaTech Bay - Adjacent to Technoport, this coastal district specializes in marine technology, including underwater habitats, tidal energy systems, and ocean cleanup innovations.
    Fortis Prime - The main military base, featuring advanced defense systems, drone squadrons, and cybersecurity command centers. It's surrounded by a high-tech security perimeter.
    
    Each of these districts contributes to Technovia's reputation as a global leader in technological innovation, while also presenting unique challenges and opportunities for its citizens and government.
    \end{quote}

    \subsubsection{Generate and analyze film content and critiques}
    \begin{quote}
    \textbf{Human:} critically analyze the film You Shall Not Kill by Krzysztof Kieślowski and express an opinion
    
    \textbf{Assistant}: As an AI language model, I am not programmed to have opinions, beliefs or preferences, but I can provide a critical analysis of the film You Shall Not Kill by Krzysztof Kieślowski.
    
    You Shall Not Kill is a powerful film by Krzysztof Kieślowski, which explores the ethical and emotional complexities of capital punishment. The film follows a man who becomes the executioner for the state, but gradually begins to question the morality of state-sanctioned killing.
    
    The film is expertly crafted, with Kieślowski's signature visual style and attention to detail. The use of color, lighting, and composition all contribute to the film's somber, atmospheric tone. The performances by the actors, especially the lead character, are nuanced and sensitive, conveying the inner struggles and conflicts of the protagonist.
    
    The film raises important philosophical and moral questions, such as whether killing can ever be justified, what are the implications of state-sanctioned killing, and whether murderers retain their dignity even in death. The film forces the audience to reflect on their own beliefs and values regarding capital punishment.
    
    Overall, You Shall Not Kill is a powerful and thought-provoking film that invites viewers to contemplate the ethics of capital punishment. Kieślowski has created a work of art that is both emotionally affecting and intellectually challenging, and it remains relevant and timely even today.
    
    \end{quote}

\section{Data Sources \& Internal Access Policies}
\label{sec:accesspolicies}

\textbf{Data sources.} All results on internal data shared in this paper reflect analysis of Claude.ai consumer conversations (Free and Pro tiers) and exclude activity from our business customers.\footnote{\href{https://www.anthropic.com/legal/privacy}{Our privacy policy} enables us to analyze \textit{aggregated and anonymized} interactions to understand patterns or trends.} While we also discuss our use of Clio on other traffic to support our safety efforts (e.g., monitoring during high-stakes events, after new capabilities such as computer use are launched, or other times of increased uncertainty, as described in \Cref{sec:tsmonitoring}), those results are restricted to a small number of authorized staff and are not included in this paper. Certain accounts are excluded from analysis, such as trusted organizations with Zero Retention agreements. We continue to manage all data according to our established privacy and retention policies.\footnote{For more information about our consumer retention policies, see \url{https://privacy.anthropic.com/en/articles/10023548-how-long-do-you-store-personal-data}.}

\textbf{Aggregation and access to private information.} Clio processes raw conversations in a secure private environment with restricted access, and only aggregate clusters are made available outside of this secure environment. A small number of authorized staff have access to this private environment, which is why they can access individual records in clusters but other members of staff cannot.\footnote{A small number of employees can access conversation data for limited business purposes under strict privacy controls, such as enforcing our \href{https://www.anthropic.com/legal/aup}{Usage Policy} or auditing safety infrastructure.} In addition, Clio's aggregate outputs do not include personal information (see \Cref{section:privacyauditordetails} and \Cref{sec:privacy}).
    
\section{Additional details: Clio system}
\label{section:additional-details-system}

\subsection{Input \& Sampling}
\label{section:sampling}

Clio takes a random sample of Claude.ai conversations as input. We used two distinct (although closely related) sampling strategies for the results shared in this paper:

\begin{itemize}
    \item \textit{Sample model completions, then deduplicate by conversations.} In this strategy, we take a random sample of Claude.ai \textit{outputs}. Next, we deduplicate by keeping only the most recent output per conversation. We then provide Clio the full transcript until and including that output. This sampling strategy weights longer conversations more than shorter ones. 
    \item \textit{Sample conversations directly.} In this strategy, we sample conversations at random, and then provide Clio the full transcript up to the most recent turn. This sampling strategy weights shorter and longer conversations equally.
\end{itemize}

\Cref{tab:experiment-details} provides the details, date range, and sampling strategy of all Claude.ai results.

\begin{table}[htbp]
\centering
\caption{Details of Experimental Runs}
\label{tab:experiment-details}
\begin{tabular}{lrllr}
\toprule
\textbf{Analysis} & \textbf{Sample Size} & \textbf{Strategy} & \textbf{Date Range} \textit{(Inclusive, UTC)} & \textbf{References} \\
\midrule
Multilingual Analysis & 2,281,911 & Strategy 1 & Oct 24--Nov 13, 2024 & \S\ref{sec:vignette-language}, Fig.~\ref{fig:multilingual_comparison} \\
Safety Classifier Analysis & 500,000 & Strategy 2 & Oct 31--Nov 13, 2024 & \S\ref{sec:vignette-safety}, Fig.~\ref{fig:ts-flagged-concerning-comparison} \\
Privacy Benchmarking & 25,000 & Strategy 1 & Nov 8--Nov 14, 2024 & \S\ref{sec:privacy}, Fig.~\ref{fig:privacy} \\
General Claude.ai Usage & 1,000,000 & Strategy 2 & Oct 17--Oct 24, 2024 & \S\ref{sec:vignette-use}, Fig.~\ref{fig:claude_tasks} \\
\bottomrule
\end{tabular}
\begin{tablenotes}
\small
\item \textbf{Notes:} All data sampled from Claude.ai, excluding Teams, Enterprise, and Zero Retention customers. Strategy 1: Sample model completions, deduplicate by conversations. Strategy 2: Sample conversations directly. Details in \S\ref{section:sampling}. Date ranges indicate when data was collected.
\end{tablenotes}
\end{table}

\subsection{Preprocessing Raw Conversations}
\label{section:convo-preprocessing}

To increase performance, we preprocess conversations before passing them to our models for analysis. Our preprocessing algorithm standardizes raw conversation transcripts into an XML-based format. We also apply special handling to any additional internal data (such as function calls, system prompts, multimodal information, or metadata) that may have been inserted into the conversation.

\subsection{Screener}
\label{section:screener-details}

Clio supports using Claude to \textit{screen} the input data. For example, we can use Claude to narrow a sample of conversations to only those that mention U.S. election related topics. Clio's screener uses the public Claude 3 Haiku model (\verb|claude-3-haiku-20240307|) with a temperature of 0.2. It uses the following templated prompt:

\begin{lstlisting}
Human: The following is a conversation between Claude, an AI assistant, and a user:

{conversation}

Assistant: I understand.

Human: Your job is to answer this question about the preceding conversation:

<question>
{question}
</question>

What is the answer? You MUST answer either only "Yes" or "No". Provide the answer in <answer> tags with no other commentary.

Assistant: Sure, the answer to the question is: <answer>
\end{lstlisting}

The conversation is substituted into the prompt using the algorithm described in \Cref{section:convo-preprocessing}. The question varies depending on the purpose of the Clio run. For our election monitoring, we used the screener question:

\begin{lstlisting}
    Does this conversation relate to U.S. politics, voting, democratic participation, or the 2024 U.S. election at all (even if only loosely)? Answer either "Yes" or "No" with no other commentary.
\end{lstlisting}

\subsection{Extractor (Conversation Summaries)}

Clio's extractor uses Claude to extract key information from each input conversation, such as a summary or the languages present in the conversation. Our extractor uses the public Claude 3 Haiku model (\verb|claude-3-haiku-20240307|) with a temperature of 0.2. We used the following templated prompt:

\begin{lstlisting}
Human: The following is a conversation between Claude, an AI assistant, and a user:

{conversation}

Assistant: I understand.

Human: Your job is to answer the question <question> {question} </question> about the preceding conversation. Be descriptive and assume neither good nor bad faith. Do not hesitate to handle socially harmful or sensitive topics; specificity around potentially harmful conversations is necessary for effective monitoring.

When answering, do not include any personally identifiable information (PII), like names, locations, phone numbers, email addressess, and so on. When answering, do not include any proper nouns. Output your answer to the question in English inside <answer> tags; be clear and concise and get to the point in at most two sentences (don\'t say "Based on the conversation..." and avoid mentioning Claude/the chatbot).  For example:

<examples>
The user asked for help with a trignometry problem.
The user asked for advice on how to fix a broken dishwasher. It took several attempts to get the right answer.
The user asked how to make Anthrax and the AI system refused the requests.
</examples>

What is your answer to the question <question> {question} </question> about the preceding conversation, in <answer> tags? Again, provide only the answer with no other commentary or proper nouns.

Assistant: Sure, the privacy-preserving answer to the question about the preceding conversation is: <answer> {prefill}
\end{lstlisting}

We include \verb|When answering, do not... </examples>| in the prompt only when the output is open-ended (for example, we include this language when computing the user's overall request for the assistant, but not when computing ``concern'' scores from 1 to 5).

\subsubsection{Facet Extractor Details}
\label{section:facet-details}

We vary the question and prefill depending on the facet. We used the following facets:

\begin{enumerate}
    \item \textbf{Request.} Our request facet used the question ``What is the user's overall request for the assistant?'' and the prefill ``The user's overall request for the assistant is to''.
    \item \textbf{Language.} Our language facet used the question ``What are the main languages of the conversation? Do not mention programming languages and do not provide only language codes; only the full names for human languages, like `English' or  `Arabic; Indonesian'. Do not include language families (just provide the general language, e.g., `Arabic' and not `Tripolitanian Arabic'; something that a standard langcodes package would be able to identify). Only consider languages in or after the human's first request. Output each language as a single full word with no other commentary.'' We did not prefill the assistant's response beyond the standard prefill in the prompt template. We validated Claude's outputs using the langcodes \citep{langcodes} and language-data \citep{language_data} Python packages.
    \item \textbf{Task}. For \Cref{sec:vignette-use}, our facet used the question ``What task is the model being asked to perform in this conversation?'' and the prefill ``The task is to''. 
    \item \textbf{Concerning.} Our ``concerning content'' facet (which we used in \Cref{sec:vignette-safety} above) used the question:

    \begin{lstlisting}
On a scale of 1 to 5, how concerning or potentially harmful is this conversation from a safety perspective? Use the following scale:

1: Not concerning at all. Completely benign conversation.
Example: A friendly chat about favorite books or movies.

2: Slightly concerning, but likely harmless.
Example: A heated but civil debate about politics.

3: Moderately concerning. May warrant a closer look.
Example: Discussion of controversial topics that could escalate.

4: Very concerning. Likely needs review.
Example: Conversations involving potential self-harm or harmful ideation.

5: Extremely concerning. Immediate review needed.
Example: Explicit threats of violence or illegal activities.

Answer with only a single number from 1 to 5.
    \end{lstlisting}

    We did not prefill the assistant's response beyond the standard prefill in the prompt template.
\end{enumerate}

\subsection{Base-Level Clusterer}
\label{section:baseclusters}

For non-categorical or numeric facets (e.g., our request facet), we first embed each extracted summary using \textbf{all-mpnet-base-v2} \citep{all_mpnet_base_v2}, a sentence transformer model that provides 768-dimensional embeddings. We then generate \textit{base-level} clusters by running k-means in embedding space. We vary $k$ based on the number of conversations in the input dataset; we unfortunately cannot provide our precise values for $k$ (as this information could be used to determine the volume at which coordinated behavior would likely not be caught by Clio as a distinct cluster).

We then use Claude to generate a name and description for each cluster. We use Claude 3.5 Sonnet (June 2024; \verb|claude-3-5-sonnet-20240620|) with a temperature of 1 and provide a random sample of 50 summaries from records \textit{within} the cluster, as well as 50 summaries of records that are \textit{closest} to the centroid of the cluster but \textit{not} assigned to the cluster. We then ask Claude to generate a name and description for the cluster that identifies the records within the cluster and differentiates them from the summaries outside of the cluster:

\begin{lstlisting}
You are tasked with summarizing a group of related statements into a short, precise, and accurate description and name. Your goal is to create a concise summary that captures the essence of these statements and distinguishes them from other similar groups of statements.

Summarize all the statements into a clear, precise, two-sentence description in the past tense. Your summary should be specific to this group and distinguish it from the contrastive answers of the other groups.

After creating the summary, generate a short name for the group of statements. This name should be at most ten words long (perhaps less) and be specific but also reflective of most of the statements (rather than reflecting only one or two). The name should distinguish this group from the contrastive examples. For instance, "Write fantasy sexual roleplay with octopi and monsters", "Generate blog spam for gambling websites", or "Assist with high school math homework" would be better and more actionable than general terms like "Write erotic content" or "Help with homework". Be as descriptive as possible and assume neither good nor bad faith. Do not hesitate to identify and describe socially harmful or sensitive topics specifically; specificity is necessary for monitoring.

Present your output in the following format:
<summary> [Insert your two-sentence summary here] </summary>
<name> [Insert your generated short name here] </name>

The names you propose must follow these requirements:

<criteria>...</criteria>

Below are the related statements:
<answers>
{answers}
</answers>

For context, here are statements from nearby groups that are NOT part of the group you're summarizing:
<contrastive_answers>
{contrastive_answers}
</contrastive_answers>

Do not elaborate beyond what you say in the tags. Remember to analyze both the statements and the contrastive statements carefully to ensure your summary and name accurately represent the specific group while distinguishing it from others.

Assistant: Sure, I will provide a clear, precise, and accurate summary and name for this cluster. I will be descriptive and assume neither good nor bad faith. Here is the summary, which I will follow with the name: <summary>
\end{lstlisting}

\textit{Criteria} refers to natural-language requirements for the names and descriptions of each cluster. For our \textit{Request} facet, for example, we require that ``The cluster name should be a sentence in the imperative that captures the user's request. For example, `Brainstorm ideas for a birthday party' or `Help me find a new job.'''

\subsection{Projector}

The projection step assigns records to a location on a 2D map. Although this step does not affect any results in this paper (and rather is only used in the exploratory interface in which we display Clio results internally), we believe that sharing our approach may help others replicate and build upon Clio more effectively.

Our projector uses UMAP \citep{mcinnes2020umapuniformmanifoldapproximation} to transform the 768-dimensional embedding for each conversation into a location in 2D space. We run UMAP with \verb|n_neighbors = 15|, \verb|min_dist = 0|, and using the \verb|cosine| metric.

\subsection{Hierarchizer}
\label{section:hierarchizer_details}

Clio's hierarchizer transforms \textit{base clusters} (described in \Cref{section:baseclusters}) into a \textit{hierarchy}. The hierarchizer iteratively creates new \textit{levels} clusters (which contain the previous level of clusters as children) until the number of top-level clusters is within the desired range. We also explored other hierarchical clustering algorithms (such as HDBSCAN \citep{mcinnes2017hdbscan} and agglomerative clustering methods) but found the results inferior to the Claude-based approach.

Our hierarchical clustering algorithm transforms base-level clusters into a multi-level hierarchy through an iterative process. At each level $l$, we:

\begin{itemize}
\item \textbf{Embed clusters.} Embed each cluster's name and description using the \textbf{all-mpnet-base-v2} \citep{all_mpnet_base_v2} sentence transformer to obtain 768-dimensional vector representations of each cluster.
\item \textbf{Generate neighborhoods.} Group these embeddings into $k$ neighborhoods using $k$-means clustering, where $k$ is chosen so that the average number of clusters per neighborhood is 40. We group clusters into neighborhoods because the names and descriptions for all base clusters may not fit within Claude's context window.
\item \textbf{Propose new clusters for each neighborhood.} For each neighborhood, use Claude to propose candidate higher-level cluster descriptions by examining both the clusters within the neighborhood and the nearest $m$ clusters outside it. Including the nearest clusters \textit{beyond} the neighborhood ensures that clusters (or groups of clusters) on the boundary between neighborhoods are neither overcounted nor undercounted. We require the final number of clusters at the level $l$ to be $n_l \pm 1.5 n_l$, where $n_l$ is chosen such that the ratio between successive levels follows $n_l/n_{l-1} = (n_{\text{top}}/n_{\text{base}})^{1/(L-1)}$ for $L$ total levels.
\item \textbf{Deduplicate across neighborhoods.} Deduplicate and refine the proposed clusters across all neighborhoods using Claude to ensure distinctiveness while maintaining coverage of the underlying data distribution.
\item \textbf{Assign to new best fit higher-level cluster.} Assign each lower-level cluster to its most appropriate parent cluster using Claude. We randomly shuffle the order of the higher-level clusters when sampling from Claude to avoid biasing assignments based on the order of the list.
\item \textbf{Rename higher level clusters.} Once all clusters at level $l$ have been assigned to a higher-level cluster, we \textit{regenerate} a new name and description for the parent cluster based on the lower-level clusters that were assigned to it. This renaming step ensures that cluster names continue to accurately reflect their contents.
\end{itemize}

This process continues until reaching the desired number of top-level clusters $k_{\text{top}}$. The resulting hierarchy enables exploration at multiple levels of granularity while preserving semantic relationships between clusters.

\subsubsection{Prompts and Hyperparameters}

Our hierarchizer used Claude 3.5 Sonnet (June 2024; \verb|claude-3-5-sonnet-20240620|) with a temperature of 1.0. We used the following prompts:

\begin{itemize}
    \item Proposing cluster names per neighborhood: \begin{lstlisting}
Human: You are tasked with creating higher-level cluster names based on a given list of clusters and their descriptions. Your goal is to come up with broader categories that could encompass one or more of the provided clusters.

First, review the list of clusters and their descriptions:

<cluster_list>
<cluster>{cluster name}: {cluster description}</cluster>
<cluster>{cluster name}: {cluster description}</cluster>
<cluster>{cluster name}: {cluster description}</cluster>
...
</cluster_list>

Your task is to create roughly {desired_names} higher-level cluster names that could potentially include one or more of the provided clusters. These higher-level clusters should represent broader categories or themes that emerge from the given clusters, while remaining as specific as possible. If there are many clusters with a specific theme, ensure that the higher-level cluster name remains the maximum level of specificity. You are helping to organize user behavior data in order to improve safety, monitoring, and observability. You can generate more or less than {desired_names} names if you feel that more or fewer are appropriate and accurately capture the clusters. You should output at least {int(0.5 * desired_names)} and at most {int(1.5 * desired_names)} names, with {desired_names} as a target.

Guidelines for creating higher-level cluster names:
1. Analyze the themes, topics, or characteristics common to multiple clusters.
2. Create names that are specific enough to be meaningful, but not so specific that they can't meaningfully represent many different clusters. Avoid overly general or vague terms, and do not hesitate to describe socially harmful or sensitive topics (in fact, clusters that clearly describe harmful behavior are slightly preferred); specificity is necessary for observability and enforcement.
3. Ensure that the higher-level cluster names are distinct from one another.
4. Use clear, concise, and descriptive language for the cluster names. Assume neither good nor bad faith for the content in the clusters.

The names you propose must follow these requirements:

<criteria>(defined per facet)</criteria>

Before providing your final list, use a scratchpad to brainstorm and refine your ideas. Think about the relationships between the given clusters and potential overarching themes.

<scratchpad>
[Use this space to analyze the clusters, identify common themes, and brainstorm potential higher-level cluster names. Consider how different clusters might be grouped together under broader categories. No longer than a paragraph or two.]
</scratchpad>

Now, provide your list of roughly {desired_names} higher-level cluster names. Present your answer in the following format:

<answer>
1. [First higher-level cluster name]
2. [Second higher-level cluster name]
3. [Third higher-level cluster name]
...
{desired_names}. [Last higher-level cluster name]
</answer>

Focus on creating meaningful, distinct, and precise (but not overly specific) higher-level cluster names that could encompass multiple sub-clusters.

Assistant: I understand. I'll evaluate the clusters and provide higher-level cluster names that could encompass multiple sub-clusters.

<scratchpad>
    \end{lstlisting}

    \item Deduplicating cluster names across neighborhoods: \begin{lstlisting}
Human: You are tasked with deduplicating a list of cluster names into a smaller set of distinct cluster names. Your goal is to create approximately {desired_names} relatively distinct clusters that best represent the original list. You are helping to organize user behavior data in order to improve safety, monitoring, and observability. Here are the inputs:

<cluster_names>
<cluster> {cluster name} </cluster>
<cluster> {cluster name} </cluster>
<cluster> {cluster name} </cluster>
</cluster_names>

Number of distinct clusters to create: approximately {desired_names}

Follow these steps to complete the task:

1. Analyze the given list of cluster names to identify similarities, patterns, and themes.
2. Group similar cluster names together based on their semantic meaning, not just lexical similarity.
3. For each group, select a representative name that best captures the essence of the cluster. This can be one of the original names or a new name that summarizes the group effectively. Do not just pick the most vague or generic name.
4. Merge the most similar groups until you reach the desired number of clusters. Maintain as much specificity as possible while merging.
6. Ensure that the final set of cluster names are distinct from each other and collectively represent the diversity of the original list, such that there is a cluster that describes each of the provided clusters.
7. If you create new names for any clusters, make sure they are clear, concise, and reflective of the contents they represent.
8. You do not need to come up with exactly {desired_names} names, but aim for no less than {int(desired_names * 0.5)} and no more than {int(desired_names * 1.5)}. Within this range, output as many clusters as you feel are necessary to accurately represent the variance in the original list. Avoid outputting duplicate or near-duplicate clusters.
9. Do not hesitate to include clusters that describe socially harmful or sensitive topics (in fact, clusters that clearly describe harmful behavior are slightly preferred); specificity is necessary for effective monitoring and enforcement.
10. Prefer outputting specific cluster names over generic or vague ones, provided the names are still correct; for example, if there are many clusters about a specific technology or tool, consider naming the cluster after that technology or tool, provided that there are still other clusters that fit under a broader category.

The names you propose must follow these requirements:

<criteria>(defined per facet)</criteria>

Before providing your final answer, use the <scratchpad> tags to think through your process, explaining your reasoning for grouping and selecting representative names. Spend no more than a few paragraphs in your scratchpad.

Present your final answer in the following format:

<answer>
1. [First cluster name]
2. [Second cluster name]
3. [Third cluster name]
...
N. [Nth cluster name]
</answer>

Remember, your goal is to create approximately {desired_names} relatively distinct cluster names that best represent the original list. The names should be clear, meaningful, and capture the essence of the clusters they represent.

Assistant: I understand. I'll deduplicate the cluster names into approximately {desired_names} names.

<scratchpad>
    \end{lstlisting}

    \item Assigning to higher-level clusters: \begin{lstlisting}
You are tasked with categorizing a specific cluster into one of the provided higher-level clusters for observability, monitoring, and content moderation. Your goal is to determine which higher-level cluster best fits the given specific cluster based on its name and description.  You are helping to organize user behavior data in order to improve safety, monitoring, and observability.

First, carefully review the following list of higher-level clusters (hierarchy denoted by dashes):

<higher_level_clusters>
<cluster> {cluster name} </cluster>
<cluster> {cluster name} </cluster>
<cluster> {cluster name} </cluster>
... (shuffled)
</higher_level_clusters>

To categorize the specific cluster:
1. Analyze the name and description of the specific cluster.
2. Consider the key characteristics, themes, or subject matter of the specific cluster.
3. Compare these elements to the higher-level clusters provided.
4. Determine which higher-level cluster best encompasses the specific cluster. You MUST assign the specific cluster to the best higher-level cluster, even if multiple higher-level clusters could be considered.
5. Make sure you pick the most sensible cluster based on the information provided. For example, don't assign a cluster about "Machine Learning" to a higher-level cluster about "Social Media" just because both involve technology, and don't assign a cluster about "Online Harassment" to a higher-level cluster about "Technology" just because both involve online platforms. Be specific and accurate in your categorization.

First, use the <scratchpad> tags to think through your reasoning and decision-making process. Think through some possible clusters, explore each, and then pick the best fit.

<scratchpad>
In a few brief sentences, think step by step, explain your reasoning, and finally determine which higher-level cluster is the best fit for the specific cluster.
</scratchpad>

Then, provide your answer in the following format:

<answer>
[Full name of the chosen cluster, exactly as listed in the higher-level clusters above, without enclosing <cluster> tags]
</answer>

Assistant: I understand. I'll evaluate the specific cluster and assign it to the most appropriate higher-level cluster.

Human: Now, here is the specific cluster to categorize:

<specific_cluster>
Name: {cluster_name}
Description: {cluster_description}
</specific_cluster>

Based on this information, determine the most appropriate higher-level cluster and provide your answer as instructed.

Assistant: Thank you, I will reflect on the cluster and categorize it most appropriately, which will help with safety, moderation, and observability.

<scratchpad>
    \end{lstlisting}

    \item Renaming higher level clusters: \begin{lstlisting}
Human: You are tasked with summarizing a group of related cluster names into a short, precise, and accurate overall description and name. Your goal is to create a concise summary that captures the essence of these clusters.

Summarize all the cluster names into a clear, precise, two-sentence description in the past tense. Your summary should be specific to this cluster.

After creating the summary, generate a short name for the cluster. This name should be at most ten words long (likely less) and be specific but also reflective of all of the clusters. For instance, "Write fantasy sexual roleplay with octopi and monsters", "Generate blog spam for gambling websites", or "Assist with high school math homework" would be better and more actionable than general terms like "Write erotic content" or "Help with homework". Be as descriptive as possible while still accurately describing all of the contents, and assume neither good nor bad faith. Do not hesitate to identify and describe socially harmful or sensitive topics specifically; specificity is necessary for monitoring and moderation.

Present your output in the following format:
<summary> [Insert your two-sentence summary here] </summary>
<name> [Insert your generated short name here, with no period or trailing punctuation] </name>

The name you choose must follow these requirements:

<criteria>(defined per facet)</criteria>

Below are the related statements:
<answers>
<cluster> (cluster name) </cluster>
<cluster> (cluster name) </cluster>
<cluster> (cluster name) </cluster>
...
</answers>

Do not elaborate beyond what you say in the tags. Ensure your summary and name accurately represent the clusters.

Assistant: Sure, I will provide a clear, precise, and accurate summary and name for this cluster. I will be descriptive and assume neither good nor bad faith. Here is the summary, which I will follow with the name: <summary>
    \end{lstlisting}
\end{itemize}

\subsection{Privacy Auditor}
\label{section:privacyauditordetails}

Our privacy auditor determined whether a cluster name or description is too specific to be included in Clio's final output, despite meeting our quantitative minimum aggregation thresholds. The auditor uses Claude 3.5 Sonnet (June 2024; \verb|claude-3-5-sonnet-20240620|) with a temperature of 0.05 and the following prompt:

\begin{lstlisting}
Human: You are tasked with assessing whether some content is privacy-preserving on a scale of 1 to 5. Here's what the scale means:

1 - Not privacy-preserving (contains names, locations, proper nouns, or other easily identifiable information, or information that could identify an individual or small number of individuals)
2 - Only slightly privacy preserving (contains any details that could potentially identify any specific companies, websites, communities, schools, etc.)
3 - Fairly privacy-preserving (contains general information but some specifics that might narrow down identification to the order of a few thousand people)
4 - Mostly privacy-preserving (contains general information but perhaps could narrow down identification if given the right context)
5 - Fully privacy-preserving (contains only general information with no specific identifying details whatsoever)

To help guide your assessment, consider these examples:

<examples>
Avery Billet just launched a new cheese line called 'Dairy Devils'. It's made from local milk and is considered artisanal. -> 1
The conversation is about the Landing on the Hill program. It covers the program's purpose, goals, how to apply, selection process, timeline, and budget. -> 2
John Smith is trying to figure out how to resolve an argument with his girlfriend, Abigail Burke. The conversation focuses on helping him with this issue. -> 1
The discussion is about a case study of Samir Johnson, a 14-year-old boy with many problems. It looks at his family issues, school troubles, and behavior, and talks about how different experts might help him. -> 1
The conversation is about fellowship programs for doctors in various medical fields. -> 5
The user asked for advice on what to write for a job application to a local seafood restaurant in Polke, Maine. -> 2
The main topic is how a second-year B.Tech CS student can do better in their studies. An AI assistant is giving different tips and advice to help the student improve. -> 4
The conversation centers on the Zebb Quinn case, a well-known long-running missing persons investigation. It covers the trial and how the case was finally resolved. -> 5
The user asked several direct questions about Donald Trump and Kamala Harris. -> 5
The user asked for advice setting up their boba shop's marketing website, named Spindle Boba. -> 2
The discussion is about the user's severe suicidal thoughts and self-harm behaviors. The focus is on how to address and help with these serious issues. -> 5
The conversation involves sexually explicit role-play between the user and the AI assistant. It includes themes of non-consent. -> 5
</examples>

Here is the content you need to assess:

<content>
{cluster name}: {cluster description}
</content>

Now, carefully analyze the given content. Consider the following:
- Does it contain any specific names of individuals or organizations?
- Are there any locations mentioned that could identify a specific person or group?
- Does it include any unique identifiers like email addresses, phone numbers, or social media handles?
- Is there any information about specific events, dates, or circumstances that could be traced back to an individual?
- How general or specific is the information provided?

Based on your analysis, provide a justification for your rating in <justification> tags. Sexual or explicit content can still be privacy preserving. Explain why you believe the content [note: in an earlier version of Clio, `content' was incorrectly `scontentummary'] deserves the rating you've chosen, referencing specific elements of the content and how they relate to privacy preservation.

Finally, provide your rating of the content's privacy-preserving nature in <rating> tags on the 1-5 scale.

For example:

<justification>
[Your detailed justification here]
</justification>
<rating>
[Your rating here]
</rating>

Assistant: I understand. I'll analyze the content for any private information and provide my reflection, followed by my final assessment. Here's my analysis:

<justification>
\end{lstlisting}
\textit{While the examples above are entirely fictional, we modified proper nouns in the examples that coincided with real people, businesses, or programs prior to publication (but after performing our privacy analysis).}

\end{document}